\documentclass[journal]{IEEEtran}

\usepackage{xcolor,colortbl,moreverb}
\usepackage{xargs}

\usepackage[pdftex]{graphicx}
\usepackage{subcaption}

\usepackage{comment}

\usepackage[bottom]{footmisc} 
\usepackage[normalem]{ulem}

\usepackage[acronym]{glossaries}
\setacronymstyle{long-short}	

\usepackage[hyphens]{url}

\newacronym{CT}{CT}{computed tomography}
\newacronym{FBP}{FBP}{filtered back projection}
\newacronym{TV}{TV}{total variation}
\newacronym{TNV}{TNV}{total nuclear variation}
\newacronym{SNR}{SNR}{signal-to-noise-ratio}
\newacronym{CNN}{CNN}{convolutional neural networks}
\newacronym{xray}{X-ray}{}
\newacronym{ART}{ART}{algebraic reconstruction technique}
\newacronym{ART-TV}{ART-TV}{algebraic reconstruction technique with the total variation as a regularizer}
\newacronym{SIRT}{SIRT}{simultaneous iterative reconstruction technique}
\newacronym{SART}{SART}{simultaneous algebraic reconstruction technique}
\newacronym{CP}{CP}{Chambolle-Pock algorithm}
\newacronym{CP-TV}{CP-TV}{Chambolle-Pock algorithm with TV}
\newacronym{GAN}{GAN}{generative adversarial network}
\newacronym{unet}{U-Net}{}
\newacronym{PCA}{PCA}{principal component analysis}
\newacronym{SSIM}{SSIM}{structural similarity index}
\newacronym{MAE}{MAE}{mean absolute error}
\newacronym{AWGN}{AWGN}{additive white Gaussian noise}

\newcommandx{\fig}[1]{Fig. #1} 
\newcommandx{\sect}[1]{Sec. #1} 

\newcommandx{\todo}[1]{{\color{red} TODO: #1}}
\newcommandx{\wail}[1]{{\color{cyan} \textit{[wamus]:} #1}}
\newcommandx{\chke}[1]{{\color{orange} \textit{[chke]:} #1}}
\newcommandx{\ullu}[1]{{\color{purple} \textit{[ullu]:} #1}}
\newcommandx{\sorgre}[1]{{\color{green} \textit{[sorgre]:} #1}}

\begin{document}

\title{Sparse-View Spectral CT Reconstruction Using Deep Learning}
\author{Wail~Mustafa,
        Christian~Kehl,
        Ulrik~Lund~Olsen,
        S{\o}ren~Kimmer~Schou~Gregersen,\\
        David~Malmgren-Hansen,
        Jan Kehres,
        and~Anders~Bjorholm~Dahl
        
\thanks{Manuscript received XXX X, 2021; revised XXX X, 2021 ; accepted XXX XX, 2021. Date of publication XXX X, 2021. This work was supported by Innovation Fund Denmark (project 10437) and the EIC FTI program (project 853720). The associate editor coordinating the review of this manuscript and approving it for publication was XXXX.
(\textit{Corresponding author: Wail Mustafa.}) }
\thanks{W. Mustafa, S. Gregersen, D. Malmgren-Hansen, and A. Dahl are with the Department of Applied Mathematics and Computer Science, Technical University of Denmark, Richard Petersens Plads 324, DK-2800 Kongens Lyngby, Denmark. (e-mail: wamus@dtu.dk; sorgre@dtu.dk; dmal@dtu.dk; abda@dtu.dk).}
\thanks{U. Olsen, and J. Kehres are with the Department of Physics, Technical University of Denmark, Fysikvej 304, DK-2800 Kongens Lyngby, Denmark. (e-mail: ullu@dtu.dk; jake@fysik.dtu.dk).}
\thanks{C. Kehl was with the Department of Applied Mathematics and Computer Science, Technical University of Denmark. He is now with the Department of Physics and the Department of Information and Computing Sciences, Utrecht University, Princetonplein 5, 3584 CC Utrecht, The Netherlands. (e-mail: c.kehl@uu.nl). }
        }

\maketitle

\begin{abstract}
   Spectral computed tomography (CT) is an emerging technology capable of providing high chemical specificity, which is crucial for many applications such as detecting threats in luggage. This type of application requires both fast and high-quality image reconstruction and is often based on sparse-view (few) projections. The conventional filtered back projection (FBP) method is fast but it produces low-quality images dominated by noise and artifacts in sparse-view CT. Iterative methods with, e.g., total variation regularizers can circumvent that but they are computationally expensive, as the computational load proportionally increases with the number of spectral channels. Instead, we propose an approach for fast reconstruction of sparse-view spectral CT data using a U-Net convolutional neural network architecture with multi-channel input and output. The network is trained to output high-quality CT images from FBP input image reconstructions. Our method is fast at run-time and because the internal convolutions are shared between the channels, the computational load increases only at the first and last layers, making it an efficient approach to process spectral data with a large number of channels. We have validated our approach using real CT scans. Our results show qualitatively and quantitatively that our approach outperforms the state-of-the-art iterative methods. Furthermore, the results indicate that the network can exploit the coupling of the channels to enhance the overall quality and robustness. 
\end{abstract}

\begin{IEEEkeywords}
Computed tomography, tomographic reconstruction, spectral CT, multi-channel CT, multi-spectral CT, few-view CT, sparse-view CT, photon-counting, spectral detectors, deep learning, U-Net.
\end{IEEEkeywords}

\section{Introduction}
\label{sec:intro}


\IEEEPARstart{S}{pectral} \gls{CT} has recently emerged following the advancement in the photon-counting \gls{xray} detection technology \cite{si2019spectral,curtis2019quantification, alvarez1976energy,henrich2009pilatus, iwanczyk2010photon}. The photon-counting detectors can resolve the energy of the received photons and, thereby, distribute the photons into discrete channels according to their energies. The \gls{CT} reconstruction of such data allows us to obtain the \gls{xray} attenuation coefficients as multi-channel energy profiles, providing higher chemical specificity compared with the energy-integrating detectors used for the conventional single-energy and dual-energy \gls{CT} \cite{jimenez2016developing,rinkel2011experimental,beldjoudi2012optimised}. 

\begin{figure}
\centering
\includegraphics[width=1.0\columnwidth]{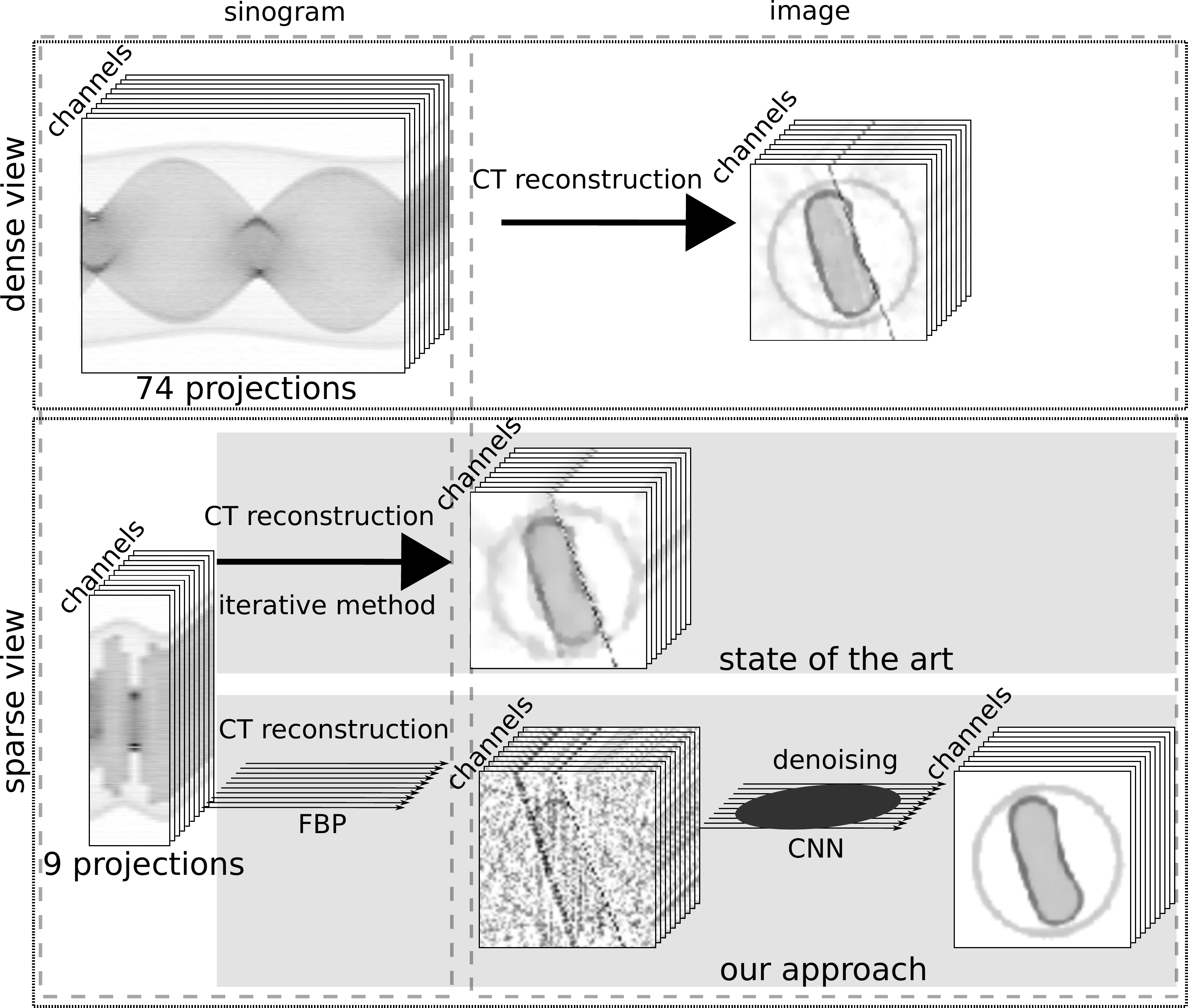}%
\caption{\label{fig:banner} Our approach to sparse-view spectral CT. The approach is faster than the state of the art and yields better reconstruction quality. }
\end{figure}

The high specificity offered by spectral \gls{CT} leads to improved performance in common tasks such as image segmentation \cite{martin2014enhanced, martin2015learning}. More importantly, high specificity is crucial in applications that require the identification of materials (e.g. detection of threats in luggage \cite{martin2014enhanced,kehl2018distinguishing,busi2019method,jumanazarov2020system}), which we pay special attention to in this work. Furthermore, resolving the full spectral shape with the spectral detectors makes it easier to tackle well-known reconstruction artifacts, namely metal streaks and beam hardening \cite{shikhaliev2005beam,shikhaliev2008energy,Nasirudin2015}.

Certain requirements such as a reduced scanning time and a simplified mechanical setup often restrict the design of \gls{CT} scanners to have a limited number of viewpoints from which \gls{xray}s are projected into the scanned object. 
Raw \gls{xray} data acquired from a single viewpoint is called a projection and \gls{CT} reconstruction from few projections is referred to as sparse-view \gls{CT} (also as few-view or compressed-sensing \gls{CT}), which is an active field of research \cite{rangayyan1985algorithms,kudo2013image, willemink2019evolution, geyer2015state}. 

For sparse-view \gls{CT}, the conventional \gls{FBP} reconstruction method is computationally fast but it produces severe noise and structural artifacts in the reconstructed images \cite{kalender2011computed}. Reconstruction methods and extensions have been introduced for solving these ill-posed inverse problems by employing different types of optimization regularization techniques \cite{pan2009commercial,beister2012iterative} or replacing the high pass filter with iteratively calculated filters unique to the specific acquisition geometry \cite{pelt2014improving}. Nowadays, iterative methods such as the \gls{ART-TV} \cite{sidky2008image} are the standard approaches for sparse-view \gls{CT}. 

One major drawback of the iterative methods is that they are computationally expensive, even for single-energy reconstruction. For spectral \gls{CT}, the computation time grows proportionally to the number of channels. If we reconstruct the channels independently with, e.g., \gls{ART-TV}, the computation time  grow linearly. Using \gls{TNV} \cite{rigie2015joint}, which is a state-of-the-art method for joint spectral reconstruction, the computation time  grow super-linearly. It is worth noting that commercially-available spectral detectors usually provide a high number of spectral channels (e.g. 128 channels by \cite{dt2019x-card} and 6400 channels by \cite{veale2018hexitec}). Using iterative methods for spectral \gls{CT} then becomes largely impractical, particularly so for time-critical applications such as luggage scanning. 

One way to reduce the time spent on reconstruction is to apply data reduction techniques such as \gls{PCA} \cite{abdi2010principal} before reconstruction to reduce the number of channels. 
However, previous work suggests that even with data reduction, a high number of channels are still needed for material identification \cite{eger2011classification, kheirabadi2017multispectral}. Moreover, previous work also suggests that post-reconstruction data reduction causes less information loss \cite{kheirabadi2017multispectral}; it is thus preferable to reconstruct all the raw channels first and reduce them afterward if necessary. In this paper, we introduce an alternative approach to spectral \gls{CT} reconstruction employing \gls{CNN} (\fig\ref{fig:banner}).

\begin{figure}[!ht]
\centering
\subcaptionbox{}
{\includegraphics[width=0.5\textwidth]{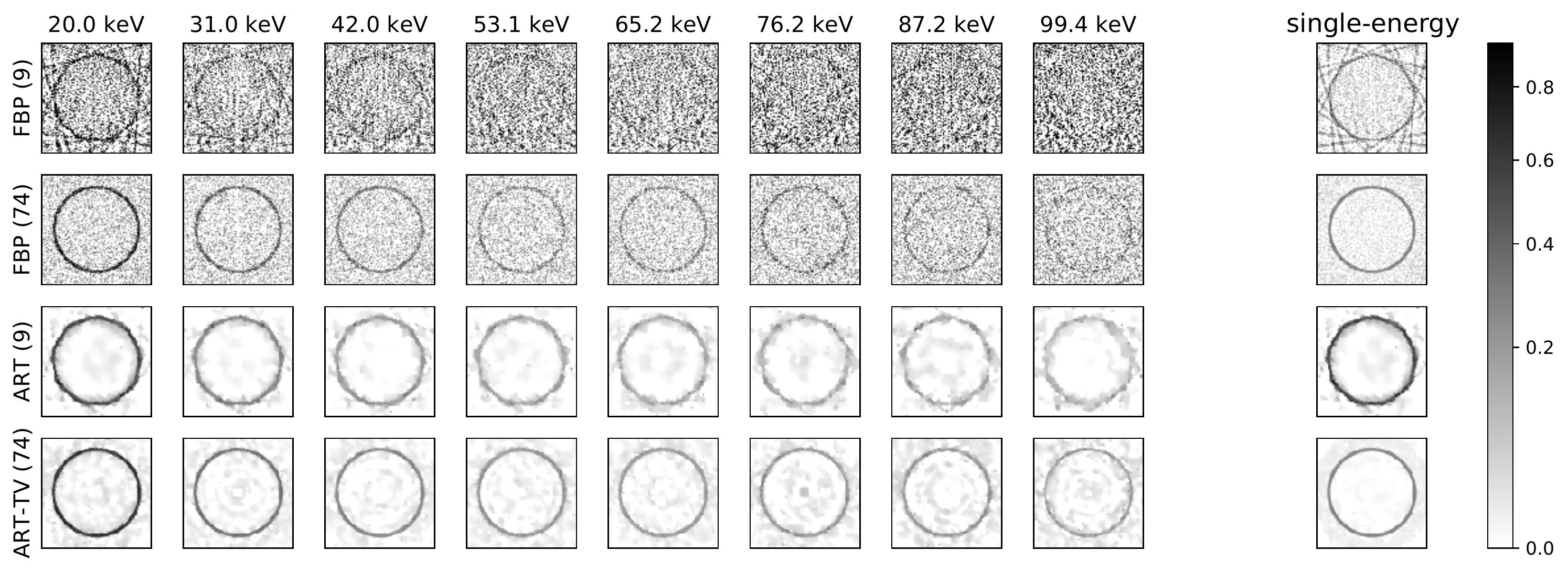}}

\subcaptionbox{}[.36\linewidth]
{\includegraphics[width=1\linewidth]{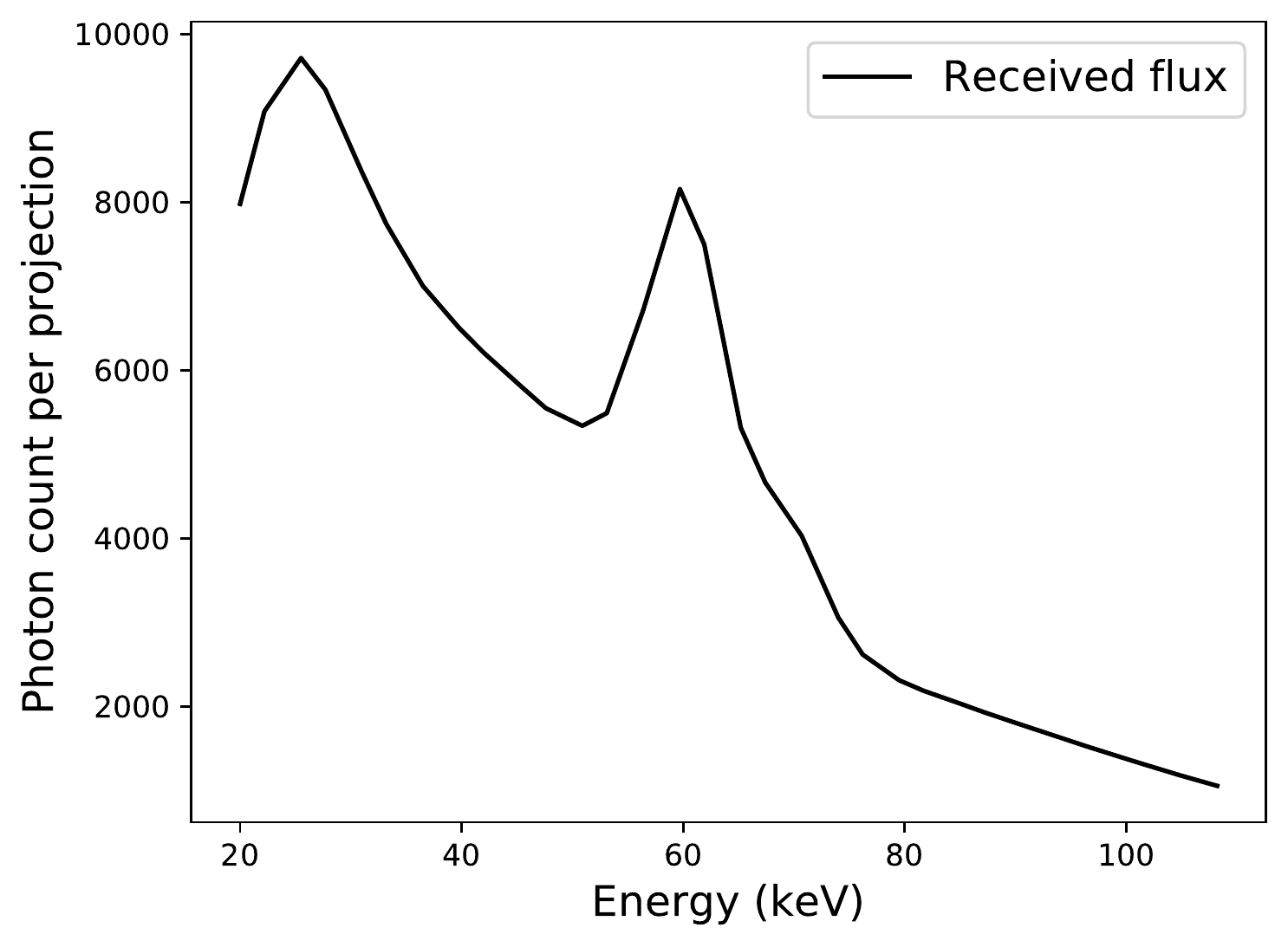} }
\subcaptionbox{}[.62\linewidth]
{\includegraphics[width=1\linewidth]{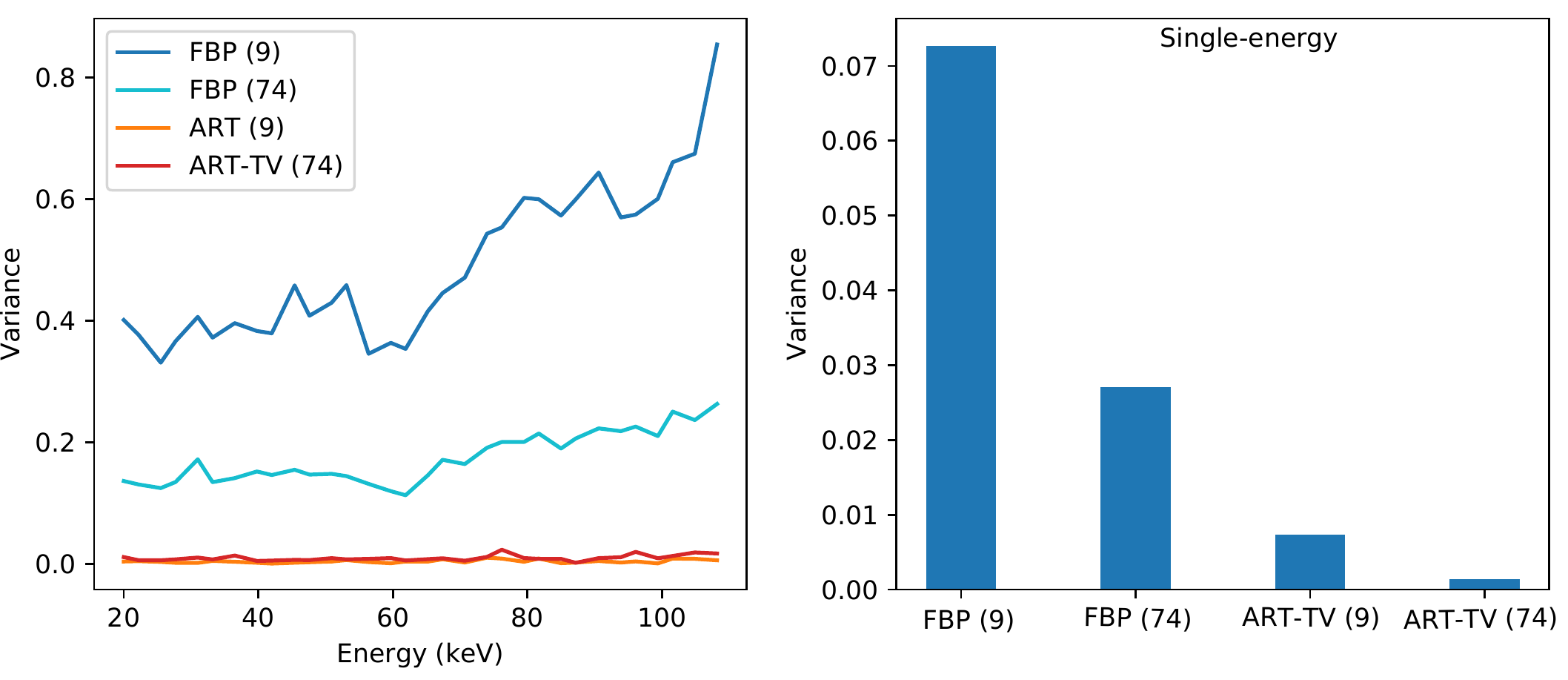}}

\caption{\label{fig:spectrum_variance} Spectral CT vs single-energy CT. In (a), the reconstructed images of the individual energy channels are shown alongside the reconstruction of their mean (approximating the behavior of a single-energy detector) for FBP and ART-TV (iterative method) with 9 (sparse-view) and 74 projections (dense-view). (b) shows how the received photons are distributed in the channels. The variance of the empty area inside the circle (which relates to the SNR) is shown in (c). Note that the variance increases as the photon count decreases (most visible in FBP (9)). In single-energy reconstruction, all photons are involved and thus we get a significantly lower variance. FBP is fast but it produces noisy images and artifacts. ART-TV circumvents the low SNR better but it is computationally expensive. }
\end{figure}

In spectral reconstruction, we obtain significantly lower \gls{SNR} per channel as compared to single-energy reconstruction with a similar acquisition setup (\fig\ref{fig:spectrum_variance}). This is because the spectral detectors distribute the received photons, which influence the reconstruction \gls{SNR}, into multiple channels whereas single-energy detectors integrate the photons in one channel. Moreover, with respect to \gls{SNR}, sparse-view \gls{CT} makes the problem more challenging compared with dense-view \gls{CT}. Imposing image priors such as \gls{TV} when using iterative reconstruction mitigates the lower \gls{SNR}. This issue can be further mitigated by a joint spectral reconstruction that exploits inter-channel correlations to maintain chemical properties such as spectral smoothness. 

In this paper, we present an approach for spectral \gls{CT} reconstruction using \gls{CNN}. The approach employs the \gls{unet} architecture \cite{ronneberger2015u} adapted to multi-channel images. As input, the network takes the spectral image reconstructed by \gls{FBP} on a channel-by-channel basis and processes the channels jointly with shared convolutional layers. The proposed approach improves the reconstruction quality, seen as piece-wise smoothness with clearly preserved edges in both the spatial and spectral domains. The reconstructed images resembles reconstructions regularized by \gls{TV} \cite{sidky2008image}. In addition to that, structural artifacts are removed and contours become smooth. Such properties are difficult if not impossible to obtain with iterative methods but can be learned by a neural network. A major advantage of using neural networks is that the model parameters are learned off-line. At run-time, only a fast forward pass through the network is needed. Furthermore, since convolutions are shared between the channels, the run-time does not significantly increase with the number of channels.

To train the model, we use real spectral \gls{CT} data \cite{kehl2018multi} supplemented with a synthetic dataset and with data augmentation during training. The model is trained with sparse-view data of only 9 projections reconstructed using \gls{FBP}. To provide a training reference for the real data, each \gls{FBP} image is paired with an image reconstructed with a high number of projections using \gls{ART-TV}. In the results, we show both qualitatively and quantitatively that our model outperforms iterative methods with \gls{TV} regularizers, which represent the state-of-the-art. Furthermore, we show that the model intrinsically handles metal artifacts, a common issue in \gls{CT} that often requires additional consideration \cite{zhang2017reduction}. 

\section{Related work}
\label{sect:sota}

Most conventional \gls{CT} systems employ single-channel \gls{xray} detectors. The reconstruction methods developed for such systems follow two approaches: direct (analytical) methods and iterative methods. The direct methods \cite{kak2002principles} are discretized versions of the inverse of the Radon transform \cite{ramm1996radon}. The most common direct method is the \gls{FBP} \cite{kak2002principles}, which first applies a high-pass filter and then employs the Fourier slice theorem to obtain the inverse. In this work, the \gls{FBP} is applied as an initial step to reconstruct each channel separately. 

The other approach employs iterative methods that are, in essence, optimization routines for solving the typically ill-posed inverse problem of \gls{CT} reconstruction. The optimization typically minimizes a data fidelity term (i.e., the back-projection error) with potentially extra regularizing terms. Based on different optimization frameworks, researchers have developed a variety of reconstruction methods, most notably \gls{ART} \cite{gordon1970algebraic, andersen1989algebraic}, \gls{SART} \cite{andersen1984simultaneous}, and \gls{SIRT} \cite{van1987numerical}.
The most common regularizer is \gls{TV} minimization \cite{rudin1992nonlinear,chambolle2004algorithm}, which represents the state-of-the-art approach for sparse-view reconstruction addressed in this article. A number of methods incorporating \gls{TV} have been developed such as \gls{ART-TV} \cite{sidky2008image}, SART-TV  \cite{yu2009compressed} and \gls{CP-TV} \cite{chambolle2011first}. 

With the emergence of photon-counting detectors, there has been a need for noise-robust reconstruction methods. To this end, researchers have been extending iterative methods through various ways of coupling the spectral channels. In \cite{barber2016algorithm}, the authors developed a method based on the primal-dual algorithm with that could achieve fast convergence but it reconstructs the spectral data into a base map of few and distinct materials, e.g., bone and brain, limiting its applicability to, e.g., medical imaging. In \cite{Ducros2017} a method for reconstructing a four-channel image of five known materials was proposed. Their method also incorporates a model of the detector response and the noise. However, in addition to limiting the reconstruction to few and known materials, the robustness of these methods in sparse-view reconstruction was not demonstrated. In this respect, other methods have been developed particularly for sparse-view spectral reconstruction. For instance, Zhang et al. proposed in \cite{zhang2016spectral} an algorithm with regularizers combining image-domain (intra-image) sparsity measures (i.e., total variation) with spectral-domain (inter-image) similarity measure (i.e. spectral mean). Also, \cite{kazantsev2018joint} used joint reconstruction of the energy channels by letting structures in favored channels propagate throughout the spectral domain.

Other methods incorporates the spectrum with a single \gls{TV} metric. This metric extends the conventional scalar TV formula applied in images with different types of channel coupling. In \cite{blomgren1998color}, the authors performed the channel coupling by global pooling of \gls{TV} across channels. This form of \gls{TV} is referred to as color \gls{TV}. In \cite{holt2014total}, the authors proposed local pooling instead using the nuclear norm of the image Jacobian and hence the method is referred to as \glsreset{TNV} \gls{TNV}. The authors showed theoretically and experimentally the superiority of their approach in most aspects of performance. The \gls{TNV} was later incorporated for spectral \gls{CT} in \cite{rigie2015joint}. 

While the spectral \gls{CT} approaches discussed above satisfy the requirement of spectral reconstruction that exploits the spectral dimensions, the major limitation is the computational cost. Our approach provides an alternative that is faster and with superior quality. Another limitation of the current iterative methods is their sensitivity to metal artifacts, which we show can be handled implicitly by our approach even in extreme cases. 

Inspired by the breakthroughs that deep learning achieved in numerous tasks \cite{goodfellow2016deep,schmidhuber2015deep,jordan2015machine,lecun2015deep, maier2018deep}, researchers have attempted to employ deep learning for \gls{CT} reconstruction. Their work can be divided into two approaches. The first approach involves building an end-to-end network that reconstructs images from sinograms. An early work employing fully-connected neural networks to model the mapping from the sinogram domain to the image domain was presented in \cite{wurfl2016deep}. In their work, the authors designed two specific networks for the parallel-beam and the fan-beam scanning geometries. Recently, \gls{CNN} have been used instead of fully-connected layers as in \cite{li2019learning, xie2019deepeff, ma2019learning}. 

The second approach for CT reconstruction with deep learning is to apply neural networks in the image domain after an initial reconstruction with \gls{FBP} \cite{mccann2017convolutional,lucas2018using}. In sparse-view reconstruction, the \gls{FBP} image would be noisy and convoluted with artifacts, making the problem resemble image denoising \cite{tian2018deep, lefkimmiatis2017non,lefkimmiatis2017non,ye2018deep,zhang2019separation}. Jin et al. \cite{jin2017deep} introduced FBPConvNet, which is an architecture based on \gls{unet} \cite{ronneberger2015u}. In their paper, the authors argued that iterative methods in certain forms in fact repetitively apply convolutions and point-wise nonlinearites. Therefore \gls{CNN} can offer a data-driven alternative. Based on that more architectures have been proposed \cite{shan2019competitive,han2018deep, kang2018deep,pelt2018improving}.  In \cite{han2018framing}, other variants of the \gls{unet} were compared. In \cite{liu2019tomogan}, the authors employed the \gls{GAN} to train the \gls{unet} as a generator. Similarly in \cite {xie2019deepenc}, \gls{GAN} was used but with an autoencoder as a generator. Note that the neural network approaches so far has been addressing single-energy \gls{CT}. Our work can be viewed as an extension of the FBPConvNet model for spectral \gls{CT} with multi-channel input and output.

\section{Architecture}
\label{sec:model} 

\begin{figure*}[!ht]
\centering
\includegraphics[width=1\textwidth]{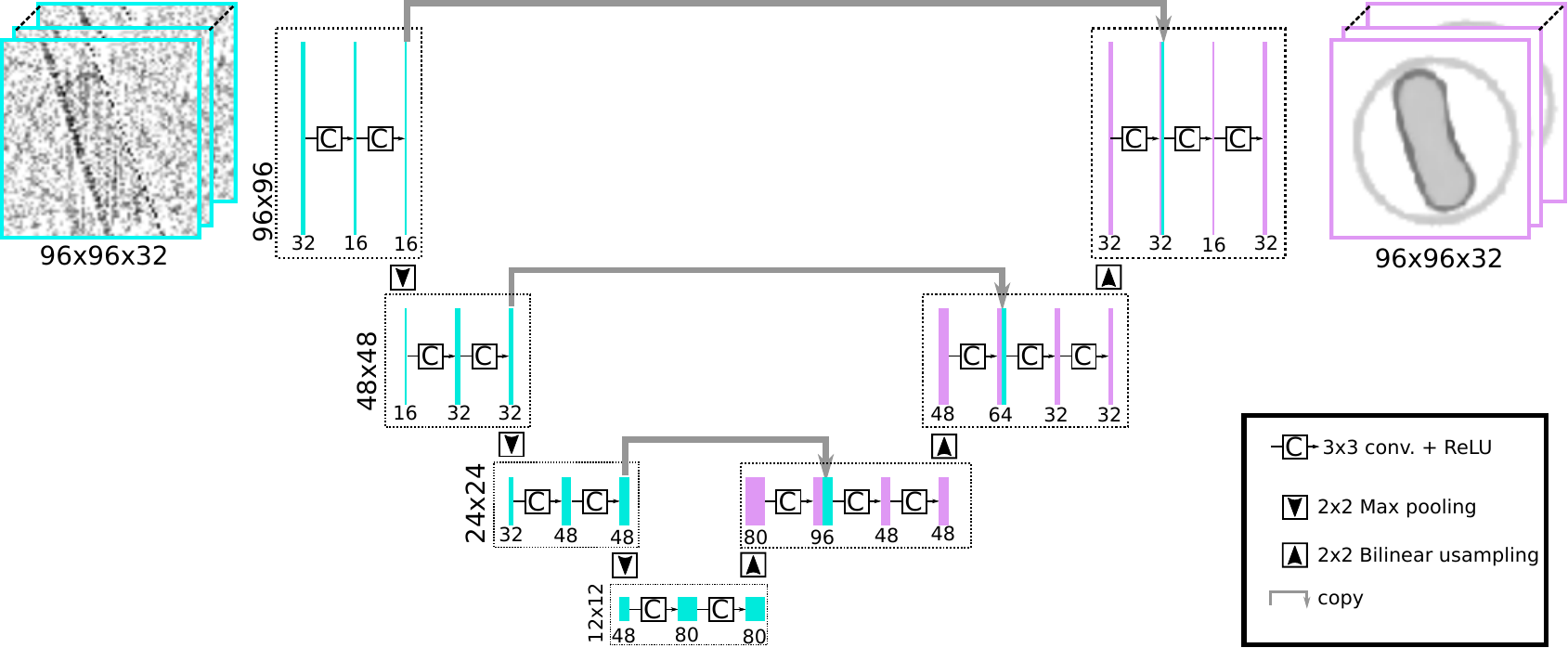}%
\caption{\label{fig:arch} The architecture of the proposed network. The network is based on the \gls{unet} \cite{ronneberger2015u} architecture. This architecture extends and modifies the one introduced in \cite{jin2017deep} with multi-channel input and output (32 channels). For better memory management, this architecture is shallower and comes with lower spatial resolution than the architectures in \cite{ronneberger2015u} and \cite{jin2017deep}. Our implementation is available at https://github.com/wailmu/spectral-ct.} 
\end{figure*}

The approach proposed in this paper consists of two steps: reconstructing each spectral channel independently using \gls{FBP} followed by refining the channels jointly using \gls{CNN} (\fig \ref{fig:banner}). We refer to this approach as Deep Spectral Inverse Radon (DSIR). The \gls{FBP} step transforms the measurement data from the sinogram domain to the image domain. As discussed in \sect \ref{sect:sota}, it would be possible to train an end-to-end network to reconstruct in one step. However, a prior reconstruction with \gls{FBP} largely simplifies the learning; the \gls{FBP} performs the domain change for which one must incorporate parameters describing the acquisition geometry of the scanner.

The architecture of the \gls{CNN} proposed in this paper is shown in \fig\ref{fig:arch}. This architecture is similar to the FBPConvNet model \cite{jin2017deep} with the extension to multi-channel input and output and with other modifications. Both models are variants of the \gls{unet} \cite{ronneberger2015u}, originally developed for image segmentation. The general \gls{unet} scheme is composed of two parts: an encoder and a decoder. The encoder gradually reduces the size of the input image using pooling while increasing the feature space by increasing the number of filters deeper in the network. This scheme allows for the derivation of higher-level features as we go down the levels in the encoder while suppressing the noise. After that, the decoder gradually up-samples the image while reducing the number of filters.

In this paper, our architecture is defined with 32 input and output channels, corresponding to the energy bins of the CT data. The multi-channel input and output convolutions provide the joint spectral processing mechanism, which is an essential element in the proposed approach. More specifically, common features are learned from all the input channels by the first convolutional layer. These common features are then propagated throughout the network and decomposed back to the original number of channels at the last convolutional layer. 

In our model, we chose to have lower spatial resolutions than the models in \cite{ronneberger2015u} and \cite{jin2017deep}. This reduction is made for practial reasons, namely, to accommodate for the memory requirements during the training phase. Specifically, the spatial resolution in our model starts with $96\times96$ at the input layer whereas in \cite{ronneberger2015u} and \cite{jin2017deep}, the resolution starts with $572\times572$ and $512\times512$, respectively. Furthermore, we made our network shallower than the ones in \cite{ronneberger2015u} and \cite{jin2017deep}. More concretely, our network contains three encoding/decoding levels as opposed to four levels in the other networks, resulting in seven instead of nine convolutional blocks.

Similar to the FBPConvNet, we apply zero-padding convolutions to obtain an output image of the same spatial resolution as the input. In the original \gls{unet}, only valid convolutions are performed, i.e., edge pixels are discarded and the resulting segmentation map has a lower resolution than the input resolution. To produce full-resolution maps, the original \gls{unet} adopts the "overlap-tile" strategy \cite{ronneberger2015u} in which the input image is processed in patches. 
The "overlap-tile" strategy used in \cite{ronneberger2015u} is logical choice for the stained biomedical cell images they use, where sub-tiles capture multiple full objects (cells). This not the case for our problem where subtiling the 96x96 pixel image would not provide the network with the full semantic view of the object we wish to reconstruct. We therefore choose to let the CNN zero-pad invalid edge pixels inside the network to produce an output of equal dimension to the input. 

Dropout regularization is applied to the last layer of each block with a dropout rate of 2\%. 
We also utilize the skip connections, which is a key element in \gls{unet}. The skip connections concatenate the features maps from the downstream layers to their corresponding layer in the decoder.
Unlike in the FBPConvNet architecture, we do not apply batch normalization \cite{ioffe2015batch}. Batch normalization acts as a regularizer and it was shown to be especially beneficial in very deep networks, e.g. the Inception architecture where the method was first applied \cite{szegedy2015going}. However, this type of regularization is undesirable in our learning problem as we aim to retrain attenuation coefficients of unbounded maximum value. This aspect is further discussed in \sect \ref{sec:arch:data_scaling}.

The network is optimized with the RMSprop optimizer \cite{hinton2012lecture} with an initial learning rate of $10^{-4}$ and a decay of $10^{-6}$ over each update. The model is trained with a batch size of 50.  The \gls{MAE} is used as loss function. We found that the \gls{MAE} leads to sharper images when compared with the mean squared error. To prevent over-fitting, augmentation is applied to the images during training before being fed to the network. The data augmentation types used here are additive white Gaussian noise and flipping of the images. 


\subsection{Data scaling} 
\label{sec:arch:data_scaling}

Besides restoring the spatial image quality, our model must retain the actual spectral profiles. Precise measurement of the spectral profiles is a key requirement for material identification~\cite{kheirabadi2017multispectral, kehl2018distinguishing}. 
Moreover, this regression problem is particularly challenging as the dynamic range of the spectrum is wide and the maximum value is unbounded---materials can have arbitrarily high attenuation coefficients. In contrast, the data in most learning tasks (e.g.,~rgb or gray-scale images) is implicitly bounded, making it straight forward to rescale the data to a certain range (typically between 0 and 1). 

In order to address the above requirements, we propose using an application-related reference value as a maximum value constraint. First, the input is scaled to this maximum value. Then, the output obtained from the network is rescaled back to the original range. By doing so, we provide the network with scaled data while retaining the actual attenuation values. For our application, we choose the reference value to be $7.66~cm^{-1}$, which is the highest attenuation coefficient of Cadmium (Cd) (i.e., at 26~keV). Although this sets a cut-off value for the attenuation coefficients, Cadmium is a highly attenuating material and all materials of interest have lower attenuation coefficients. 


\subsection{Computational complexity}
\label{sec:arch:complexity}

In order to analyze the computational complexity of our model, we need to look at the computational complexity of the \gls{FBP} part and the \gls{CNN} part. The \gls{FBP} computations are dominated by the back projection, which computes the sum of all line integrals passing through each of the reconstructed pixels. To reconstruct an $N\times N$ image from $V$ projections, the \gls{FBP} operations amounts to $O(N^2V)$ when a fixed-size discretization kernel is used (as in our case) \cite{basu2000n, jin2017deep}. To reconstruct $S$ spectral channels, the operations grow linearly, resulting in $O(SN^2V)$ operations.

There are several operations in the \gls{CNN} such as additions, upsampling, downsampling, activation functions but the computations are dominated by the convolutions. With $L$ layers, $R$ filters per layer and a $K\times K$ kernel size, the run-time evaluation of a standard \gls{CNN} requires $O(R^2 N^2 K^2 L)$ operations \cite{cong2014minimizing}. For the \gls{unet}, this complexity estimation represents an upper bound since the number of pixels are gradually reduced in the lower levels. In our network, we process the spectral channels with input and output filters similar to the internal filters. 
To accommodate for the spectral channels, we only add filters at the first and last layer; the number of internal filters are unchanged. Therefore, the added operations are still within the upper bound defined above. Furthermore, since our approach is composed of a \gls{FBP} step followed by a \gls{CNN} step, the overall complexity of our approach can be described by the operations of the step with higher complexity, i.e., the \gls{CNN}.

For the iterative methods, the computations are dominated by the back projections, which scale linearly with the number of projections and the iterations ($I$), leading to $O(SN^2VI)$ operations. The TV adds additional complexity amounting to $O(N^2T)$ operations for one channel and $O(N^2TS)$ for $S$ channels, where $T$ is the TV iterations. In the \gls{ART-TV} method, we get $O(N^2VIT)$ operations with single channel and $O(SN^2VIT)$ for $S$ channels. For \gls{TNV}, we get $O(S^2N^2VIT)$. 

The computational complexity analysis discussed above is summarized in Table \ref{fig:computation_comp} (a). When comparing methods with some parameters being method-specific, we should take into account the typical values of the parameters (Table \ref{fig:computation_comp} (b)). More specifically, note that the number of iterations ($I$ and $T$) in ART-TV are usually much higher than $K$ and $L$ in the \gls{CNN} (in our case, $I=200$ and $T=30$ whereas $K=3$ and $L=17$). This complexity analysis highlights the computational advantage of the \gls{CNN} step and thereby our model despite being an upper bound estimation. 

The computational advantage in run-time comes with higher memory requirement. With the \gls{CNN}, we need to store the pre-trained filters in all layers and the intermediate feature maps produced within a forward pass. These memory demands amounts to an upper bound of $O(S^2 N^2 K^2 L)$ \cite{rhu2016vdnn,siu2018memory}. With the current hardware technology, accepting such memory demands in exchange for fast run-time is considered a good trade-off. In the training phase however, the memory complexity increases with the batch size, i.e., the number of spectral images passed together in a single training step. This issue may limit the batch size or impose other design constraints. In our work, we chose to lower the spatial resolution as we discussed above.

\begin{table}[]
\scriptsize
\subcaptionbox{}
{
\begin{tabular}{l|l|l|}
\cline{2-3}
                          & one channel  & spectral \\ \hline
\multicolumn{1}{|l|}{FBP} &     $N^2V$        &    $SN^2V$      \\ \hline
\multicolumn{1}{|l|}{ART-TV }    &     $N^2VIT$        &   $SN^2VIT$       \\ \hline
\multicolumn{1}{|l|}{TNV}    &      --        &    $S^2N^2VIT$      \\ \hline
\multicolumn{1}{|l|}{CNN}    &    $R^2 N^2 K^2 L$         &    $R^2 N^2 K^2 L$      \\ \hline
\multicolumn{1}{|l|}{DSIR}    &     $R^2 N^2 K^2 L$        &    $R^2 N^2 K^2 L$      \\ \hline
\end{tabular}
}
\subcaptionbox{}
{
\begin{tabular}{c|c}
 parameter & typical range  \\ \hline 
 $N$ & $10^3$ \\ 
 $I$ & $10^3$ \\  
 $T$ & $10^2$ \\  
 $R$ & $10^2$\\ 
 $S$ & $10^2$\\ 
 $L$ & $10^2$\\ 
 $V$ & $10^2$\\ 
 $K$ & $10^1$\\ 
\end{tabular}
}
\caption{\label{fig:computation_comp} Computational complexity. (a) the computational complexity of our model compared with the iterative methods using the big O notation. The computational complexity of our model falls within the upper bound defined by a standard \gls{CNN}. (b) the typical ranges of the parameters in (a). Note that $N^2$ is number of pixels; $V$ is the number of projections; $I$ and $T$ are respectively the numbers of iteration for reconstruction and TV in the iterative methods; $L$, $R$ and $K^2$ are respectively the number of layers, the number of filters per layer, and the kernel size in the \gls{CNN}; and $S$ is the number of spectral channels. }
\end{table}



\section{Dataset}

\begin{figure}[!tb]
\centering
\subcaptionbox{}
{\includegraphics[width=0.7\columnwidth]{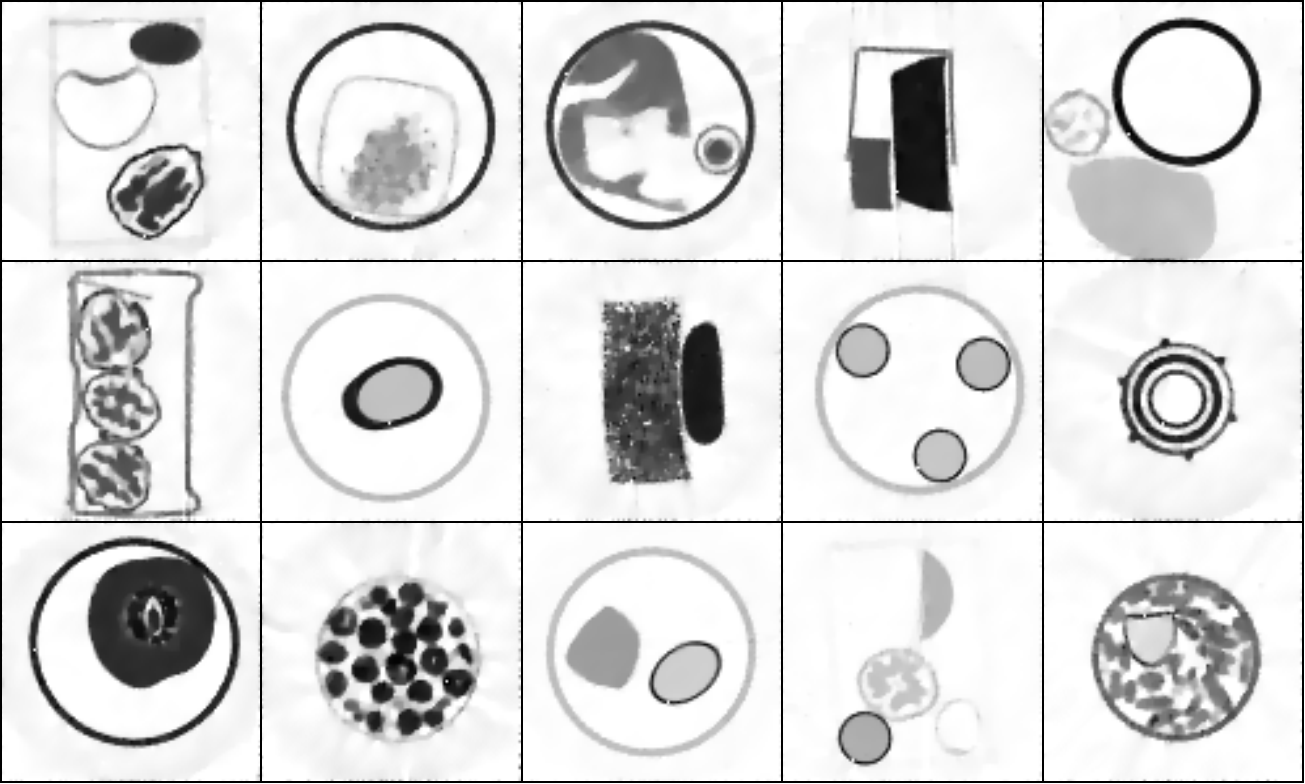}}
\subcaptionbox{}
{\includegraphics[width=0.28\columnwidth]{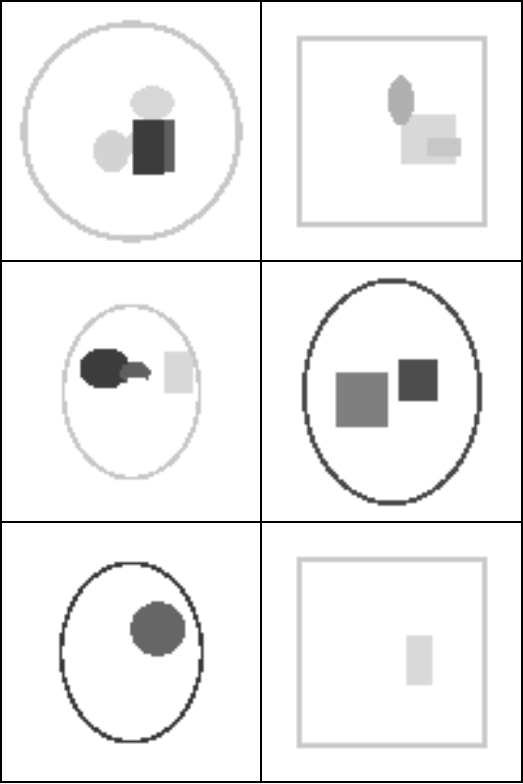}}
\caption{\label{fig:dataset_sample} Example slices from the dataset: (a) real data. (b) synthetic data. For better visualization, the images in the figure are obtained using single-energy reconstruction.}
\end{figure}

The \gls{CNN} is trained with both real and synthetic data. The data is arranged in sets of 4D volumes, each of which is composed of stacked slices of 3D spectral images. Note that the model processes each spectral image separately. In \fig\ref{fig:dataset_sample}, example slices are presented, showing the diversity of materials and geometrical shapes in the dataset. 

The real data used in this paper is composed of 22 volumes comprising in total 4350 spectral images. The data is a subset of the MUSIC dataset \cite{kehl2018multi}, which is provided by the 3D Imaging Center, DTU \footnote{The 3D Imaging Center: \url{http://imaging.dtu.dk}. The dataset is available at \url{http://easi-cil.compute.dtu.dk/index.php/datasets/music/}. }. The \gls{xray} data is acquired using the Multix Me-100-V2 detector\footnote{The new version, V3, is now sold as Detection Technology X-card ME3} and a Hamamatsu micro-focus source operated at 150~kVp and $200~\mu$A. 
Spectral detectors are subject to a number of physical effects causing deviations in the detector response. We correct the detector response using the method presented in \cite{dreier2018spectral}.

The projection data (sinograms) used in the reconstruction are computed as $\mu=-ln(I/I_0)$, which represents the ratio of photons emitted from the source ($I_0$) to photons received after being transmitted through the sample ($I$). An example of the received spectrum after correction is shown in Fig \ref{fig:spectrum_variance} (b). 
At the higher end of the spectrum, a lower number of photons are emitted, which leads to higher statistical noise. The number of detected photons are reduced further by the decreased absorption efficiency of the detector at higher energies. At the lower end of the spectrum, the sample absorbs most of the photons and very few photos are detected, leading to a high noise as well. 

The MUSIC dataset comes with 128 spectral channels covering the range between 20~keV and 160~keV. In this work however, we use 32 channels evenly distributed between between 20~keV and 108.2~keV. We chose to include the lower end of the spectrum because the material attenuation contrast is largest in this range of energies
, which are important for material identification. Due to the high levels of noise, the lower end of the spectrum thus poses a special challenge to the reconstruction methods.     

Each sample in the MUSIC dataset is acquired with 74 projections evenly distributed over $360^o$. These dense-view projection data are reconstructed with \gls{ART-TV} and used as reference images to train the \gls{CNN}. To provide the input images to the \gls{CNN}, we generate sparse-view projections from the original dense-view projections. The process is performed by sampling 9 projections from the 74 projections. Each of the sparse-view data is then reconstructed with \gls{FBP}. 

In addition to the real data, we generate synthetic data to add more geometrical variations. In this work, we generated 51 volumes containing 5100 spectral images. The synthetic volumetric objects are produced using randomly generated 3D containers made of common geometrical shapes, namely, ellipsoids, cuboids, sphubes, and elliptical cylinders. These containers are generated using TomoPhantom toolbox \cite{kazantsev2018tomophantom}. Each container is randomly assigned a material that is defined by a vector representing the spectral attenuation coefficients. The materials are selected from a database of 48 materials that were previously reconstructed and segmented from real dense-view data. 

The synthetic images are used as ground-truth references for the \gls{CNN}. To generate sparse-view synthetic data, we first produce dense-view projections through simulating the forward projection assuming the same CT acquisition geometry as the one used to generate the real data. We then, as for the real data, sample 9 evenly-spaced projections and reconstruct them with \gls{FBP}. The forward simulation and the \gls{FBP} reconstruction were performed using the Astra toolbox \cite{van2016fast}.

To train the model, the real and synthetic data are split into training and validation subsets. The training subset is composed of 13 real volumes (4350 images) and 40 synthetic volumes (4000 images), resulting in a total of 8350 spectral images. The validation subset is composed of 5 real volumes (1625 images) and 10 synthetic volumes (1000 images), resulting in a total of 2625 spectral images. Furthermore, to evaluate our approach, we keep out 4 real volumes and one synthetic volume.

\section{Results and discussion}
\label{sec:results}

\newcommand{\rulesep}{\unskip\ \vrule\ }

\begin{figure*}[!ht]
\centering
\subcaptionbox{}
{\includegraphics[width=0.493\textwidth]{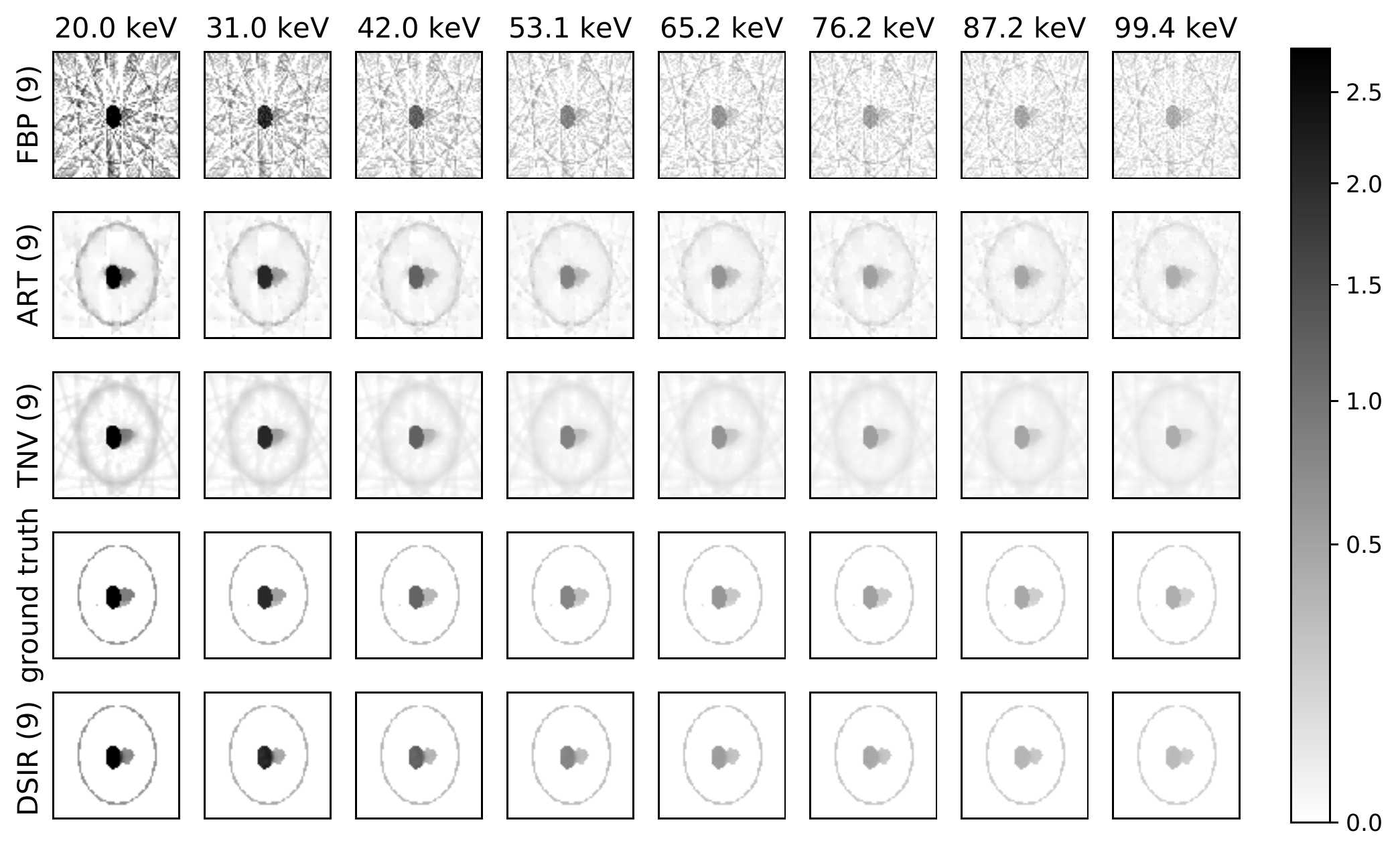}}
\rulesep
\subcaptionbox{}
{\includegraphics[width=0.493\textwidth]{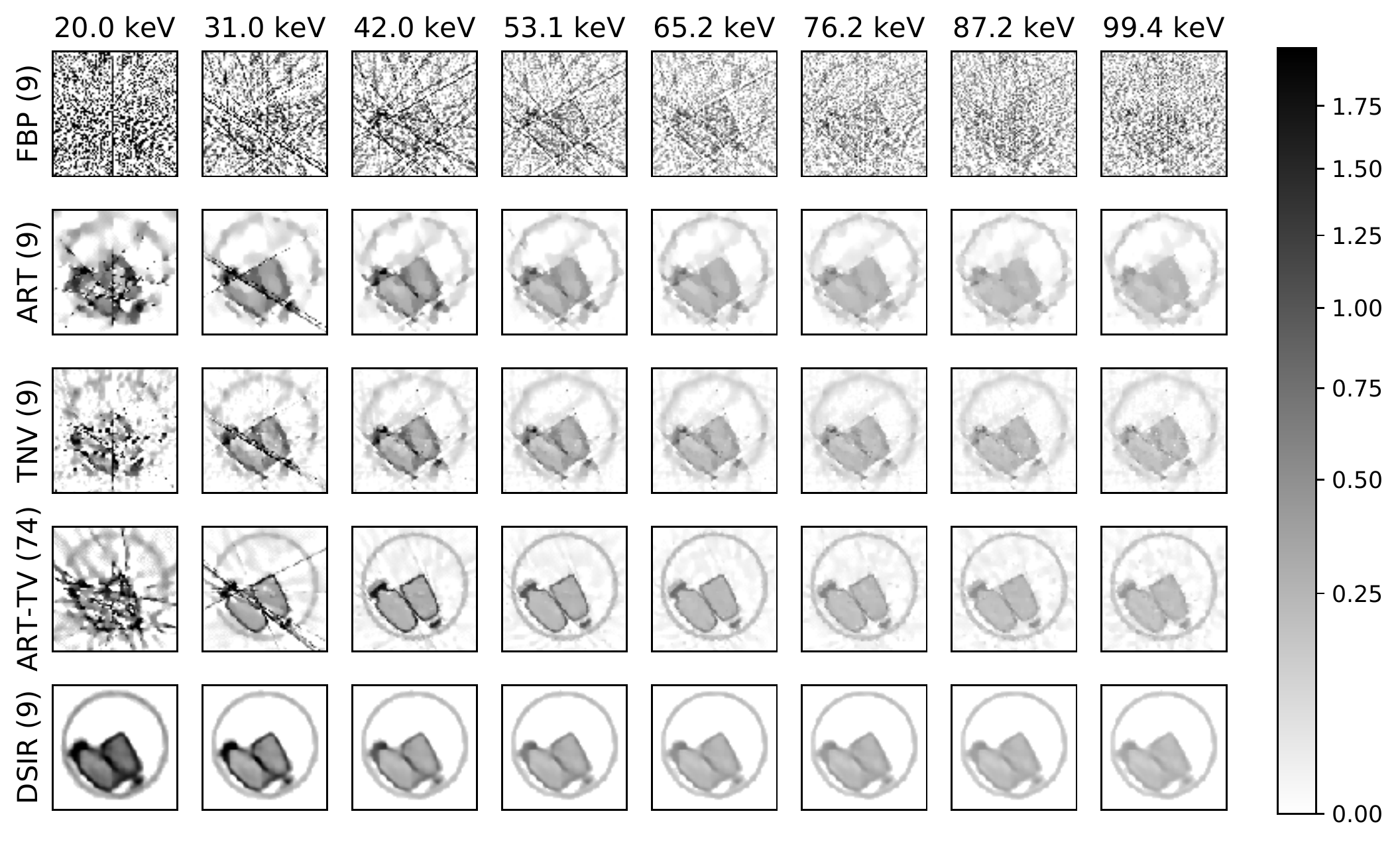}}
\subcaptionbox{}
{\includegraphics[width=0.493\textwidth]{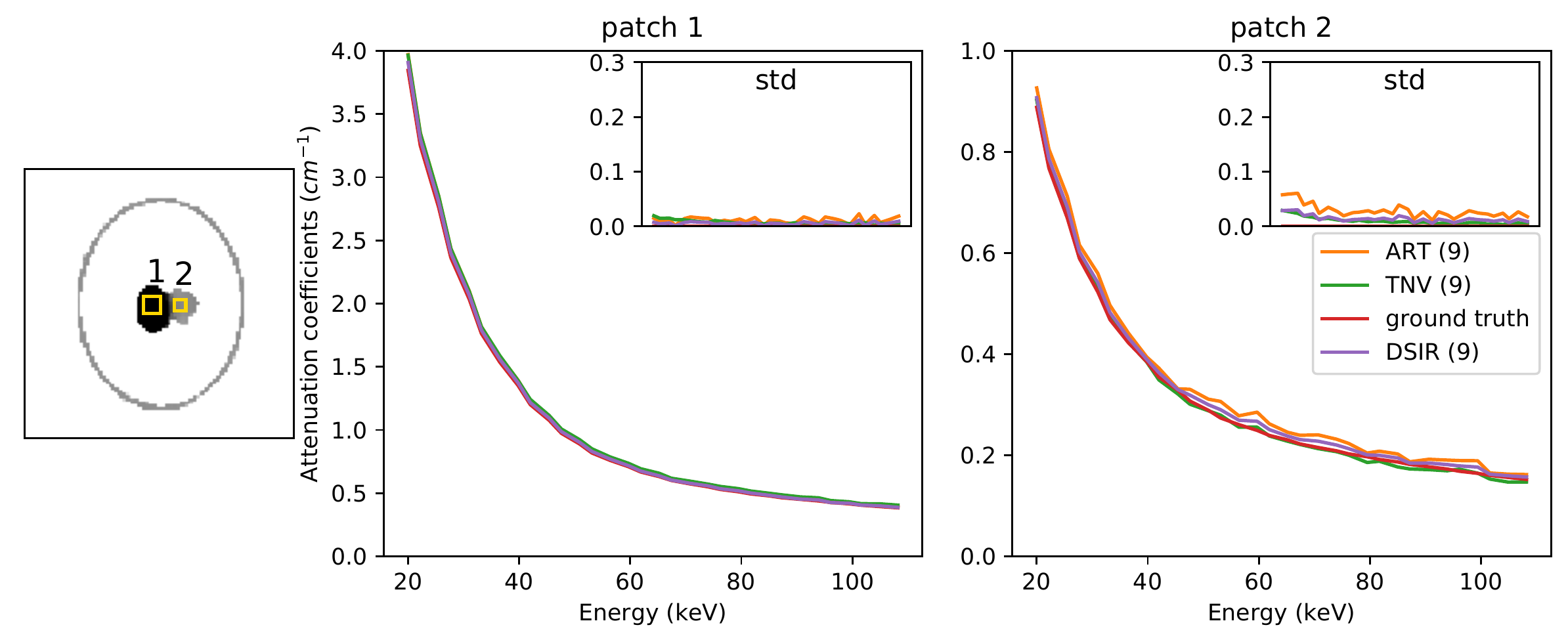}}
\rulesep
\subcaptionbox{}
{\includegraphics[width=0.493\textwidth]{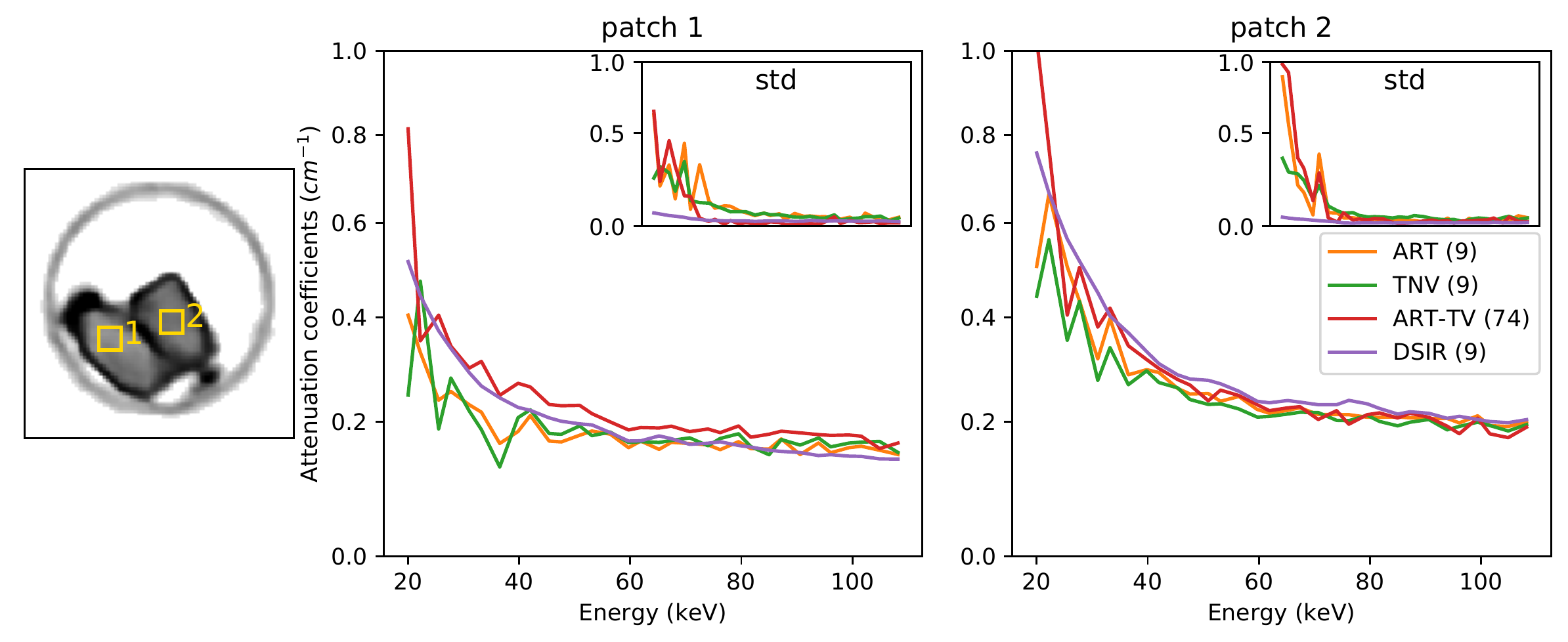}}

\subcaptionbox{}
{\includegraphics[width=0.16\textwidth]{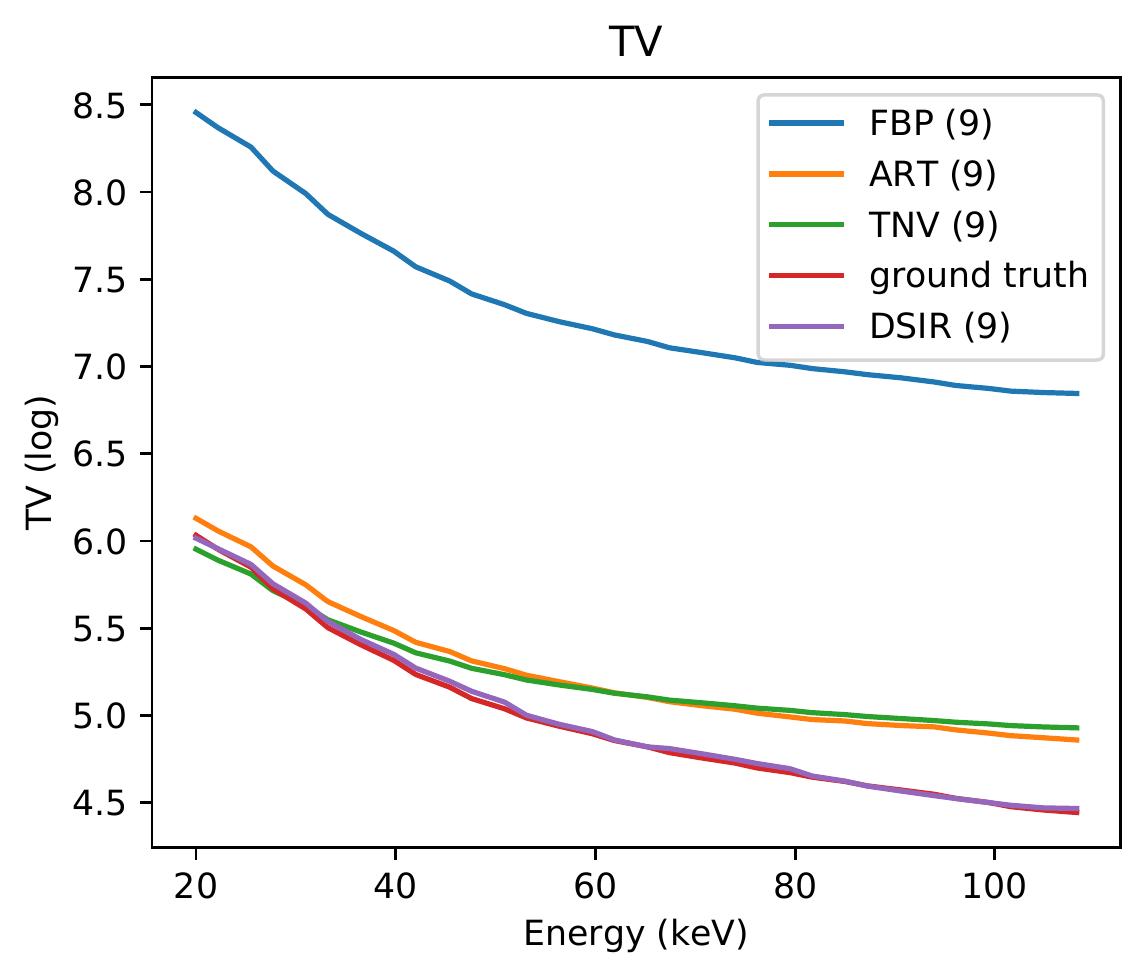}
\includegraphics[width=0.16\textwidth]{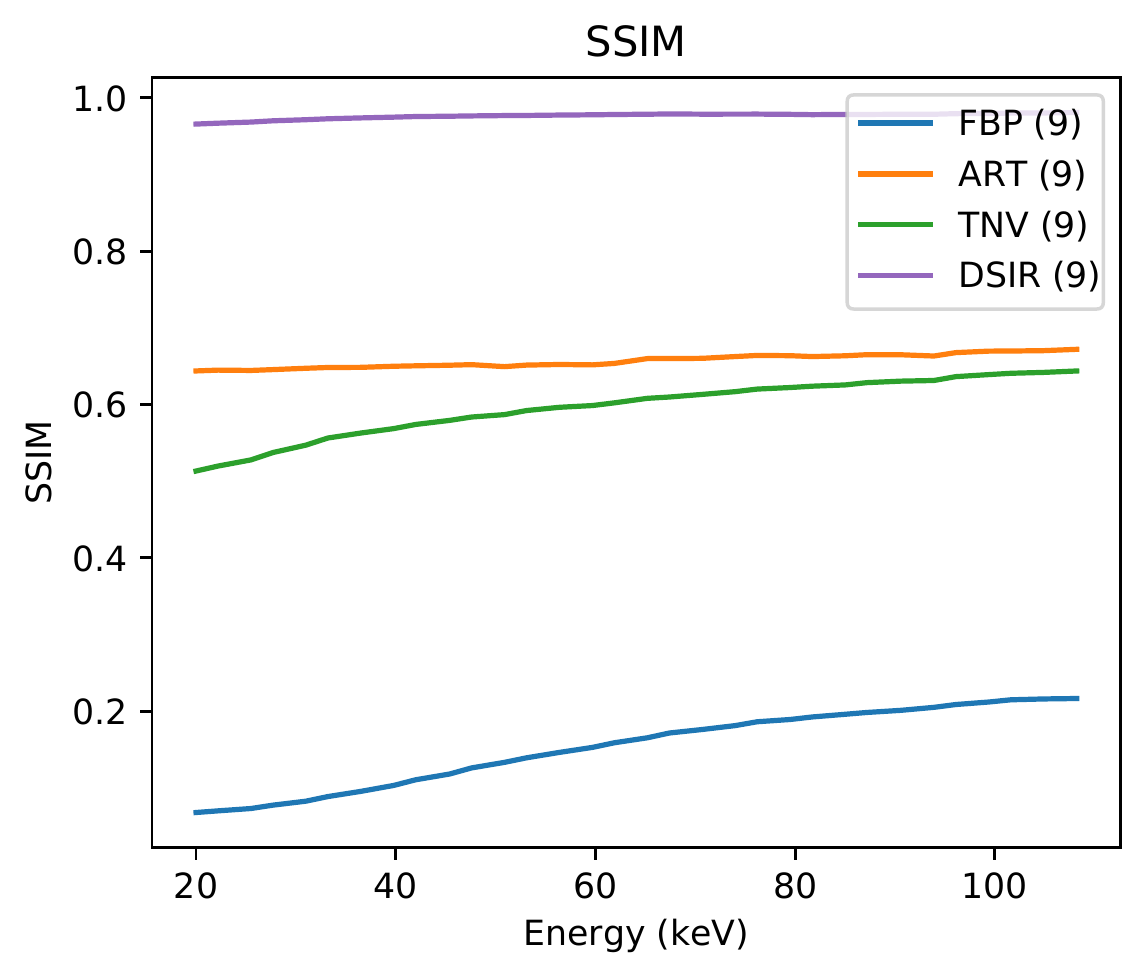}
\includegraphics[width=0.16\textwidth]{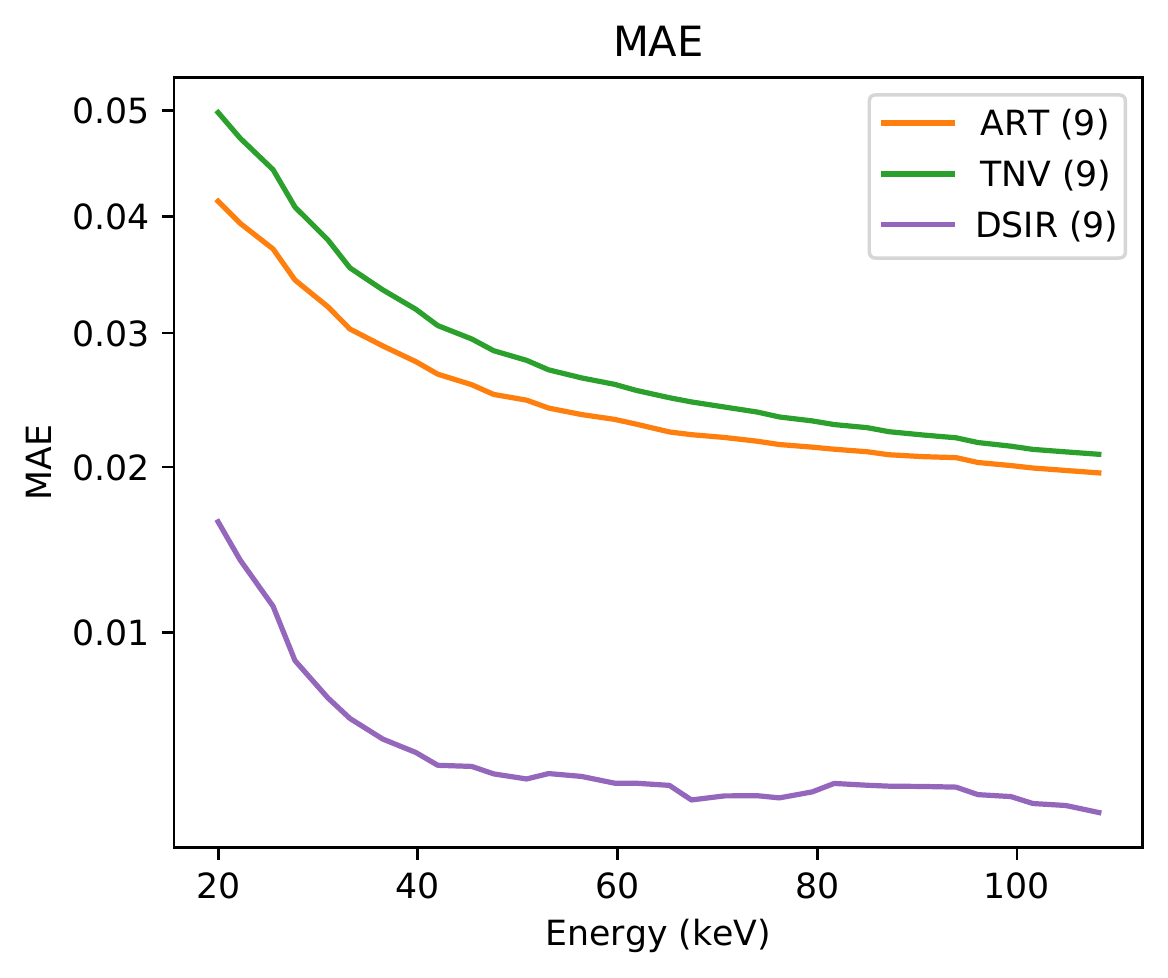}}
\rulesep
\subcaptionbox{}
{\includegraphics[width=0.16\textwidth]{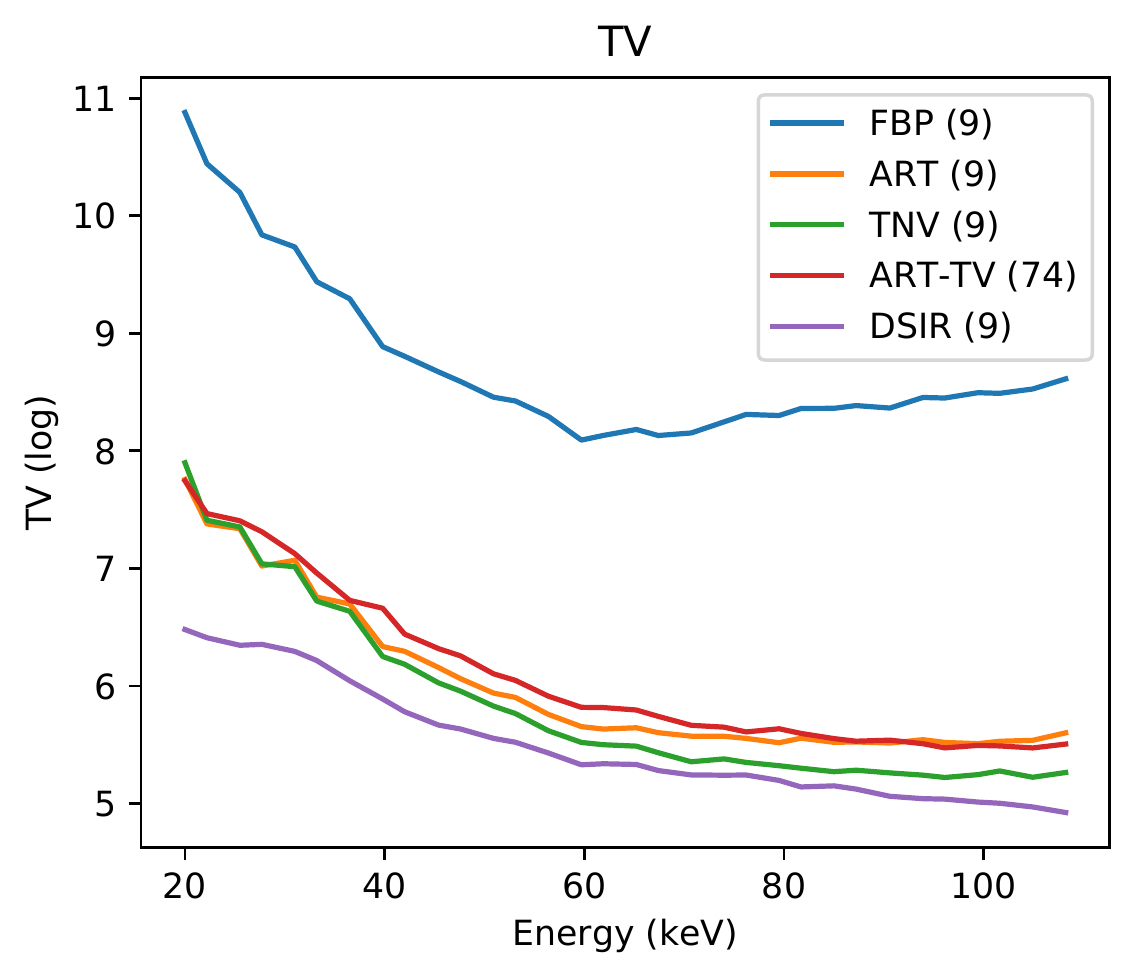}
\includegraphics[width=0.16\textwidth]{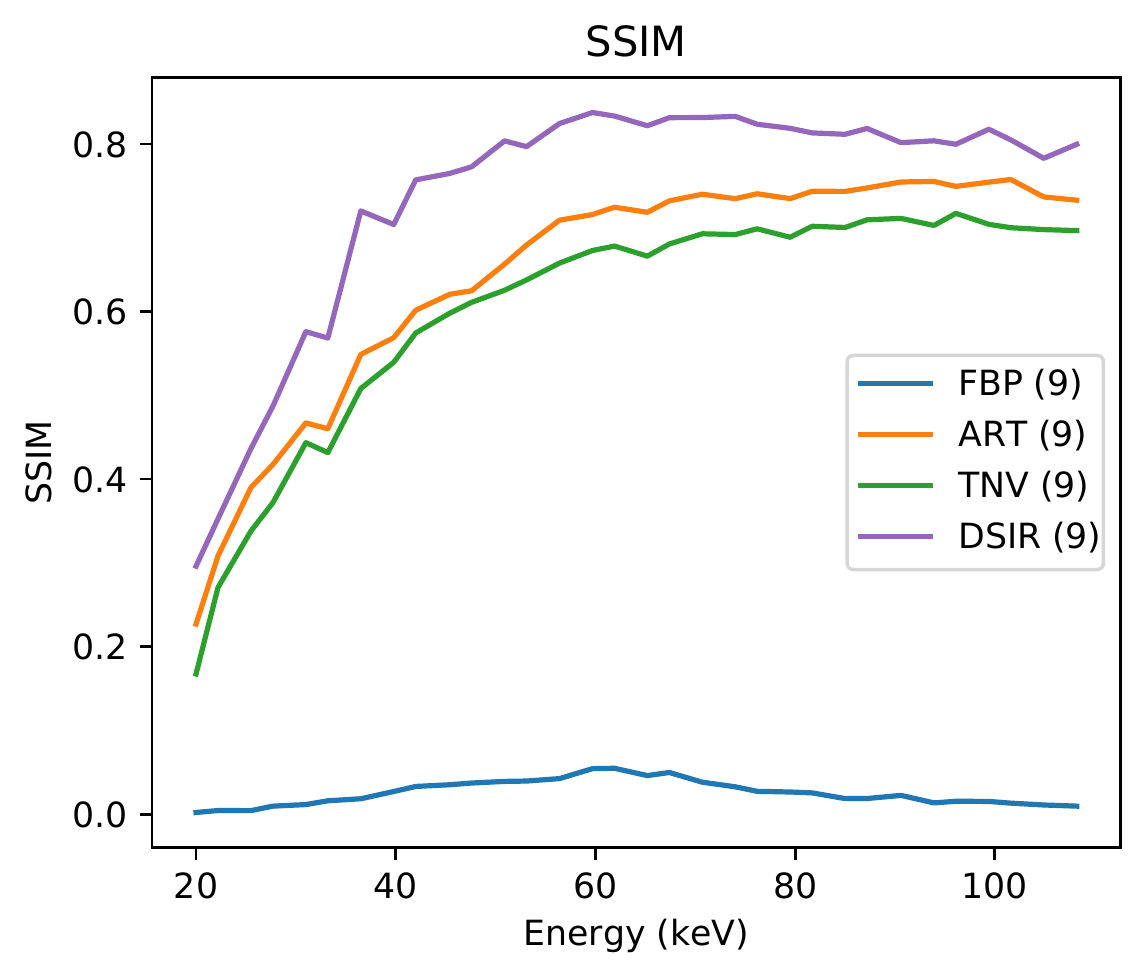}
\includegraphics[width=0.16\textwidth]{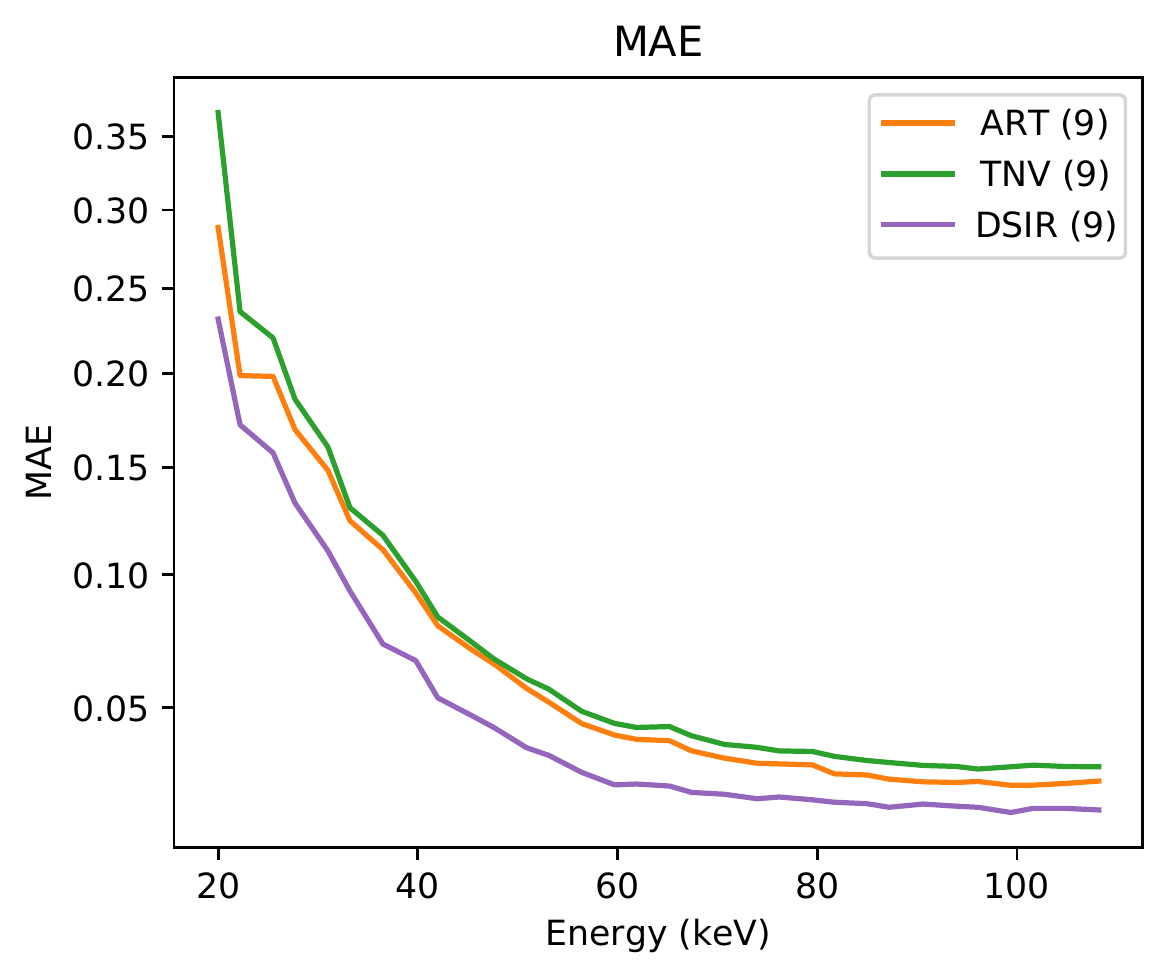}}

\caption{\label{fig:ours_vs_sot} Our approach vs iterative methods. The comparison is carried out on a synthetic slice (left) and a real slice (right). The synthetic slice comes from the synthetic test volume whereas the real slice comes from Volume I (renderings of the volumes are shown in \fig\ref{fig:volums}).  Reconstructions with other methods are compared with our approach in (a) and (b). The number between parentheses indicates the number of projections. Spectral profiles showing the mean of the attenuation coefficients and the standard deviation (std) at selected patches across 32 channels are presented in (c) and (d).  In (e) and (f), we show the image TV as well as the SSIM and the MAE metrics across all channels.}
\end{figure*}




To evaluate the model, we perform a number of experiments and comparative tests on a dedicated test subset. The objective of the experiments is to validate several aspects of the model. First, we investigate the performance properties of the proposed model on sparse-view \gls{CT} in terms of the reconstruction quality both spatially and spectrally as compared with the state-of-art approaches. Second, we would like to show that the joint spectral processing is not only faster but also enhances the reconstruction quality when compared with independent channel processing. Third, we would like to assess the robustness of joint spectral reconstruction when some channels are affected by variable noise levels. Fourth, we would like to assess the robustness to metal artifacts, a common issue in \gls{CT} imaging.

In the following, we refer to the output of our approach as 'DSIR', which is obtained after re-scaling the output of the \gls{CNN} (\sect \ref{sec:arch:data_scaling}). 
We compare our approach with \gls{ART-TV} \cite{sidky2008image} and \gls{TNV} \cite{rigie2015joint}. Moreover, we include the \gls{FBP} in the comparison; it is a standard method and also the initial step in our approach. Note that both \gls{FBP} and \gls{ART-TV} reconstruct the channels independently whereas \gls{TNV} imposes the \gls{TV} regularization spectrally. When showing real samples, we also compare with \gls{ART-TV} for reconstructions from 74 projections since that is the type of data we provide as reference to train the \gls{CNN}. 

For qualitative evaluation in the comparison below, we show images from eight spectral channels (out of 32 channels). This allows us to observe the reconstruction quality and the variation thereof across the spectrum. 
In addition to the images, we include four means of quantitative assessment:

\begin{itemize}
\item attenuation coefficient curves. The attenuation coefficient curves provides a comparative assessment indicator. We show the mean and the standard deviation of the attenuation coefficients in selected image patches belonging to different materials. The mean allows us to assess whether we can reconstruct the attenuation coefficients of materials. The mean also gives an indication for the spectral smoothness. The standard deviations allow us to compare the reconstruction quality along the spectrum. Because each patch belongs to one material, lower standard deviation indicates higher quality. 

\item \gls{TV}. Iterative methods incorporating \gls{TV} minimization represent the state of the art. We include \gls{TV} as a measure of relative image quality. Lower \gls{TV} indicates higher quality (i.e. piece-wise smoothness with edge preservation). 
It is worth noting that \gls{TV} can be misleading in some cases. For instances, an undesirably over-smoothed image would have a low \gls{TV} value. 
In the following results however, qualitative results (images) and other metrics are presented alongside the \gls{TV}.

\item \gls{MAE} and \gls{SSIM}, which are standard metrics for error quantification \cite{Zhao2017loss,Carlini2017towards,Ruder2016overview}. We show the \gls{MAE} because it is the loss function we use to train the \gls{CNN} (\sect\ref{sec:model}). The \gls{SSIM} is widely used with images as it models perceptual changes by incorporating structural information rather than absolute errors. For the real data, these metrics are computed with respect to the images reconstructed using \gls{ART-TV} with 74 projections, as no actual ground-truth images are available. Therefore, those metrics are not perfect in this case, as improvement in quality will be wrongly measured as error. Nevertheless, we use them to quantify significant improvement, which can clearly be observed qualitatively.

\end{itemize}

The evaluation aspects are presented and discussed in the following subsections:

\subsection{DSIR vs iterative methods}

We start by examining a spectral slice from the synthetic test volume and a slice from a real test volume as shown in \fig\ref{fig:ours_vs_sot}. The figure compares our approach with the iterative methods. 
The reconstructions in \fig\ref{fig:ours_vs_sot} (a) and \fig\ref{fig:ours_vs_sot} (b) show the high-quality images produced by the \gls{CNN}. Despite high noise levels in the \gls{FBP} being visible, particularly in the real sample, the \gls{CNN} is able to overcome that. The \gls{CNN} is also able to remove the reconstruction artifacts, which appear as lines in the \gls{FBP} image. 

Moreover, the \gls{CNN} also removes other types of artifacts appearing in the real slice at the lower end of the spectrum (most visible in the images at 20.0 keV and 31.0 keV). Those artifacts are due to the presence of small pieces of metal (the container's lids). 

In general, the images produced by the \gls{CNN} appear smoother and better at preserving the edges compared to the other methods with 9 projections. Compared to the images obtained with 74 projections, however, edges appear less sharp and fine shape details are lost (best viewed in \fig\ref{fig:ours_vs_sot} (b)). The loss of such details can be attributed to missing viewpoints, which the \gls{CNN} is unable to compensate for.    
On the other hand, frequently occurring features such as contours (e.g. the round container in the example) appear smooth. The reconstruction of smooth contours suggests that the \gls{CNN} is able to learn such high-level features. This property is difficult to incorporate in iterative methods.

The spectral profiles presented in \fig\ref{fig:ours_vs_sot} (c) and (d) show that we are able to obtain a significantly smoother spectrum with our approach. 
Maintaining this smoothness, even at the lower end of the spectrum, on real samples demonstrates the superiority of our approach.
The profiles also show that lower standard deviation is achieved, which confirms the high reconstruction quality seen in the images. The reconstruction quality is further confirmed in \fig\ref{fig:ours_vs_sot} (e) and (f) in which our approach consistently achieves the lowest \gls{TV} and \gls{MAE} and the highest \gls{SSIM}.

With respect to noise suppression, artifact removal, and spectral smoothness, our approach not only outperforms other methods in the sparse-view case, as clearly seen in the real sample, but it also outperforms \gls{ART-TV} with 74 projections. These results are obtained despite that the network was trained against images reconstructed with 74 projections. Such performance can be attributed to the inherent properties of \gls{unet} in noise suppression achieved by its encoder/decoder topology (\sect\ref{sec:model}). This performance also indicates that the network is able to generalize the removal of artifacts despite the artifacts being partly present in the reference data.

\subsection{Joint vs channel-by-channel reconstruction}
\label{sec:result_jointVSsingle}

\begin{figure}[!tb]
\centering
\subcaptionbox{}
{\includegraphics[width=0.50\textwidth]{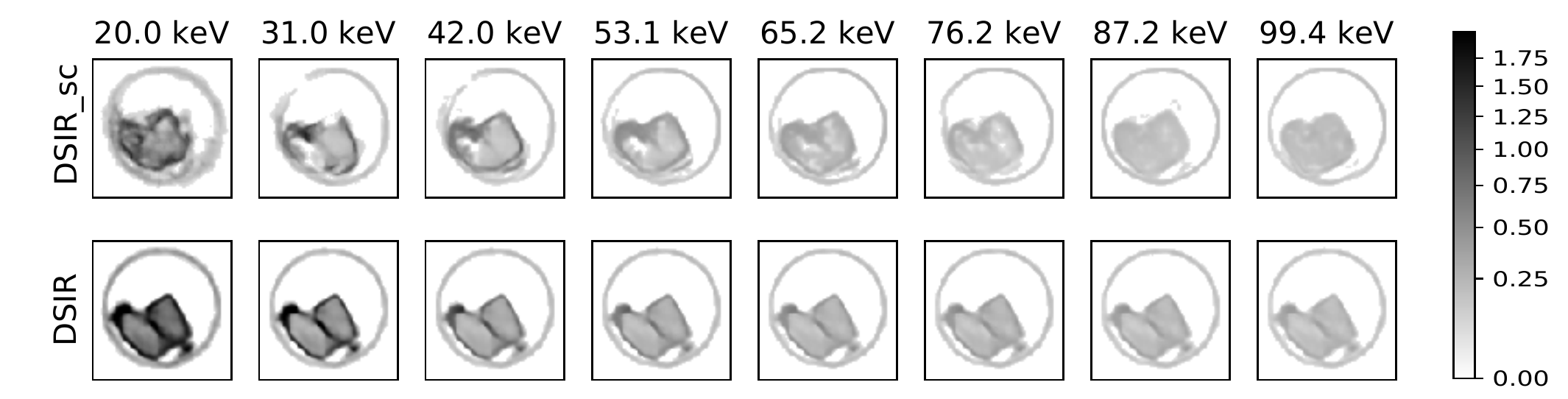}}
\subcaptionbox{}
{\includegraphics[width=0.40\textwidth]{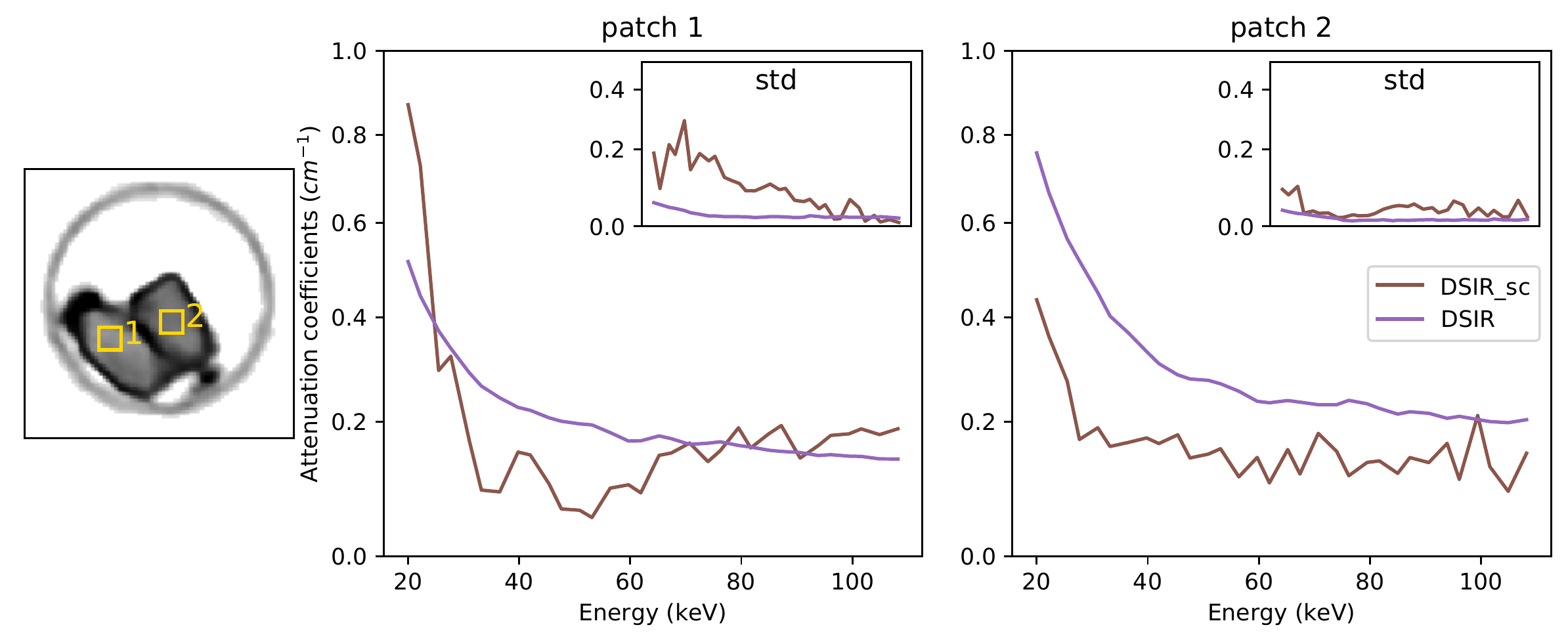}}

\caption{\label{fig:ours_mono} Joint spectral (DSIR) vs channel-by-channel (DSIR\_sc) reconstruction . In (a), reconstructions obtained when our model is trained on single-channel images are compared to our standard model. Spectral profiles from the two yellow patches are shown in (b).}
\end{figure}

The importance of joint spectral reconstruction is demonstrated in \fig\ref{fig:ours_mono} by comparing it to independent channel-by-channel reconstruction. The channel-by-channel model was realized by adapting our model to operate with single-channel input and output, i.e., similar to the model in~\cite{jin2017deep} but we apply our approach to data scaling discussed in \sect~\ref{sec:arch:data_scaling}.

The figure shows that the single-channel model is able to suppress the noise and remove the artifacts at a level comparable to our model. However, the single-channel model reconstructs shapes poorly. This indicates higher sensitivity to the low per-channel \gls{SNR}, causing inconsistent shape reconstruction in the individual channels. On the other hand, the joint convolutions appear to let our model recover the overall spectral \gls{SNR}, leading to improved shape reconstruction.

\subsection{Robustness to variable channel noise}
\begin{figure}[!ht]
\centering
\subcaptionbox{}
{\includegraphics[width=0.48\textwidth]{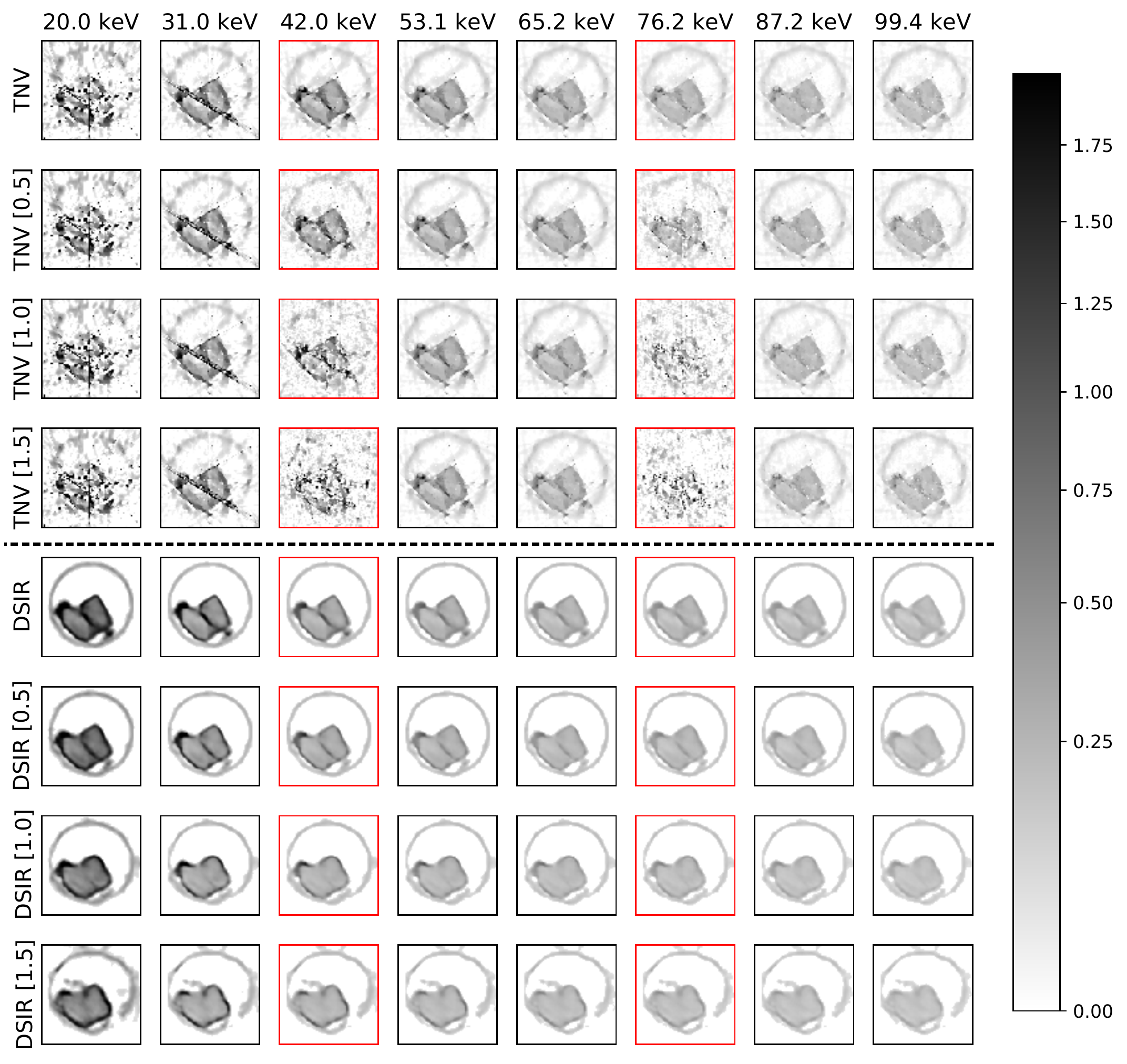}}
\subcaptionbox{}
{\includegraphics[width=0.16\textwidth]{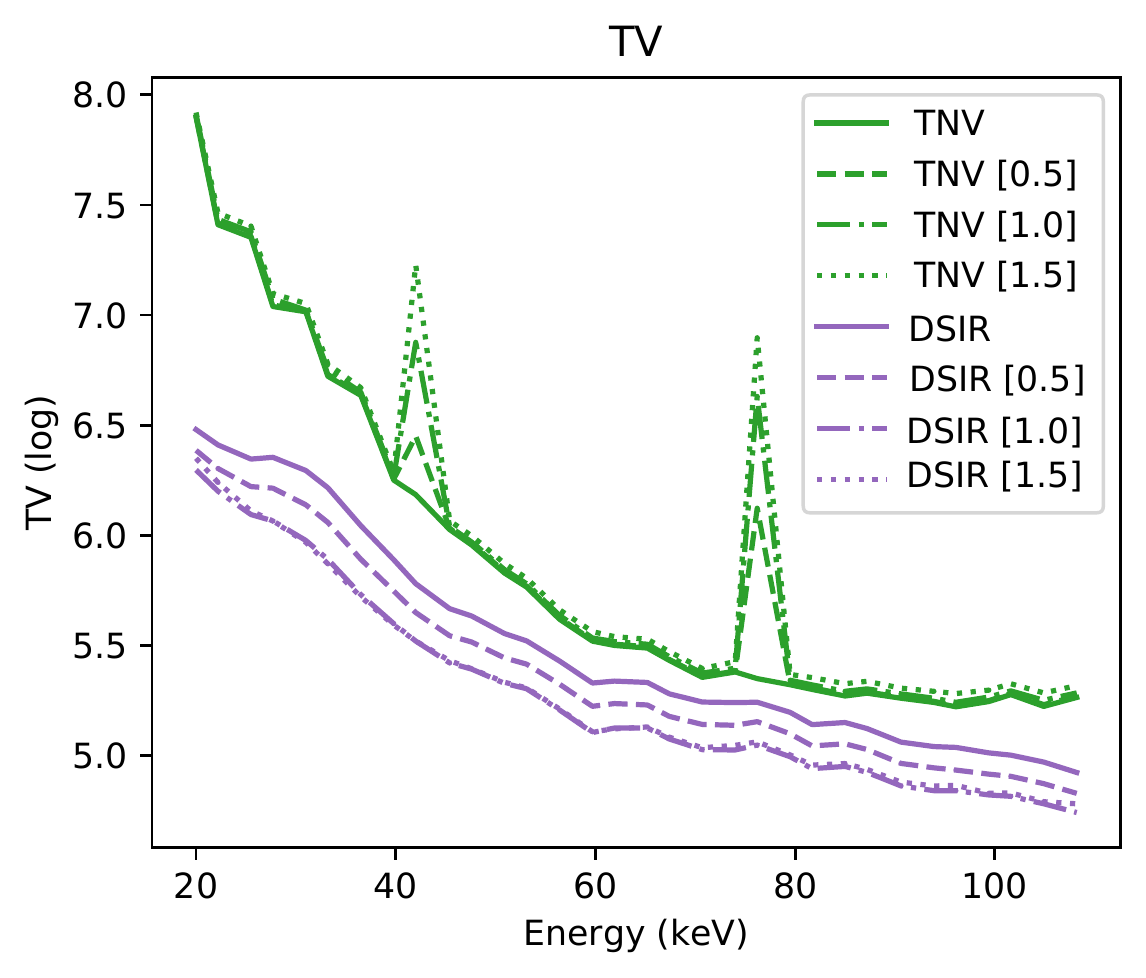}
\includegraphics[width=0.16\textwidth]{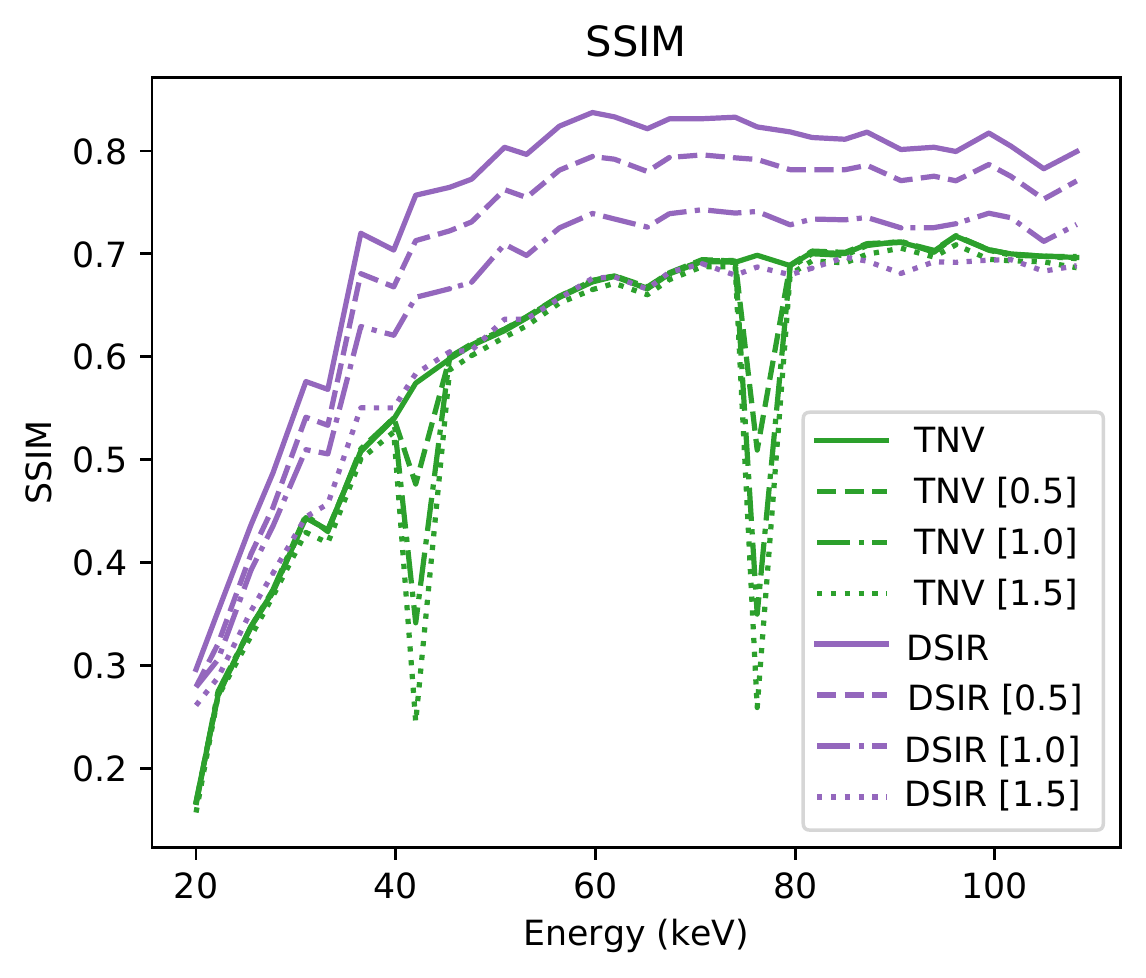}
\includegraphics[width=0.16\textwidth]{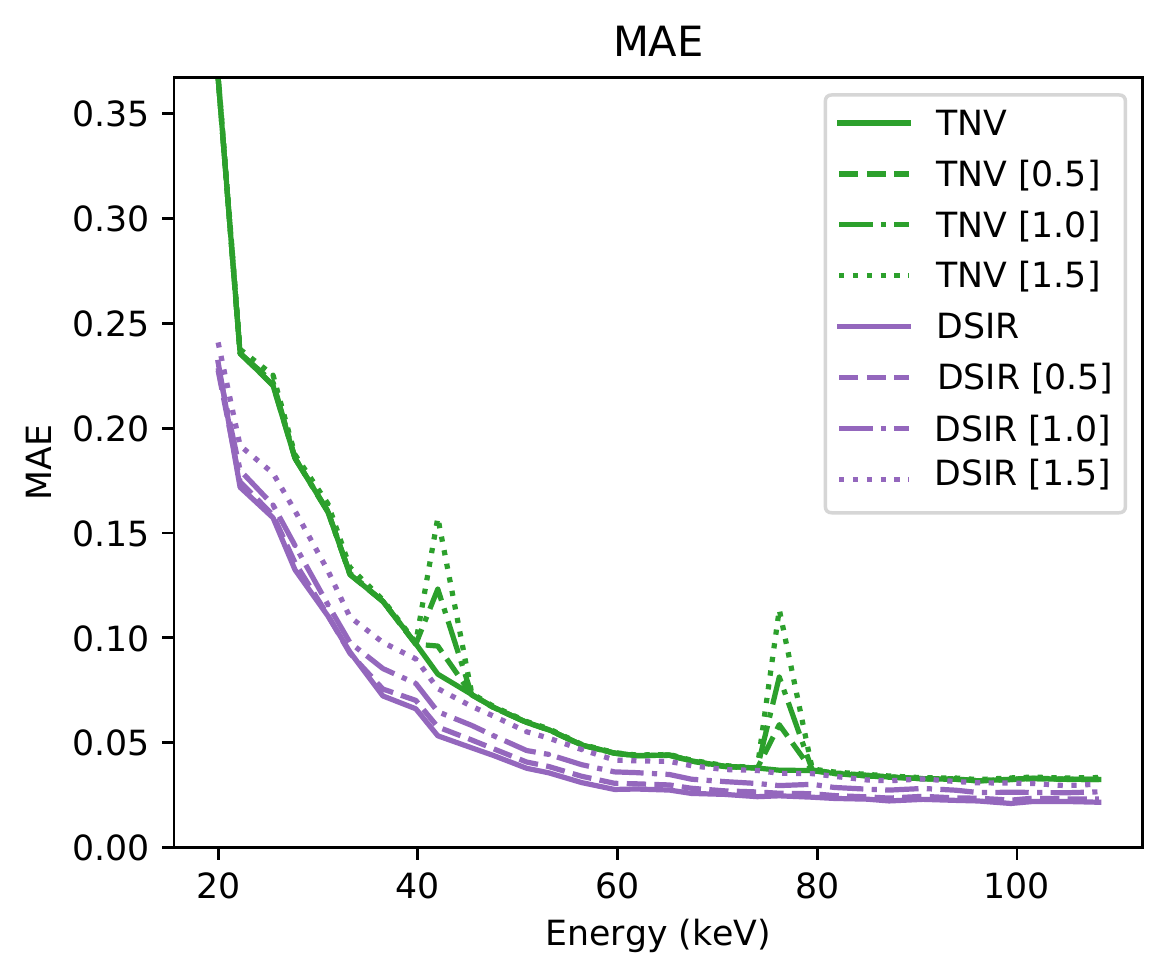}}

\caption{\label{fig:noise} The effect of adding white noise to certain channels. The numbers between parentheses indicate the standard deviation ($\sigma$) of the noise added to the channels outlined in red.  }
\end{figure}

Certain spectral channels may experience unexpected disturbances during scanning, related to the photon integration discussed in \sect \ref{sec:intro}. To assess the robustness of our model to such disturbances, we introduce \gls{AWGN} to the sinograms with a standard deviation ($\sigma$) of 0.5, 1.0 and 1.5. The noise is added to two channels: 42.0 keV and 76.2 keV. We evaluate the reconstruction quality of our model compared with \gls{TNV} as shown in \fig\ref{fig:noise}.

The figures show that our model is able to overcome a high level of noise with $\sigma=0.5~cm^{-1}$ whereas the reconstruction quality with \gls{TNV} degrades at the affected channels. However as the noise is further increased, the reconstruction quality of our model drops, resulting in distorted and over-smoothed shapes (\fig\ref{fig:noise} (a)). This degradation is clearly visible by the drop in the \gls{SSIM} (\fig\ref{fig:noise} (b)). 

The robustness to such disturbances further emphasizes the importance of joint reconstruction, as it yields a visible reconstruction quality improvement up to a certain noise level. 

\subsection{Robustness to metal artifacts}

\begin{figure}[!ht]
\centering
\subcaptionbox{}
{\includegraphics[width=0.495\textwidth]{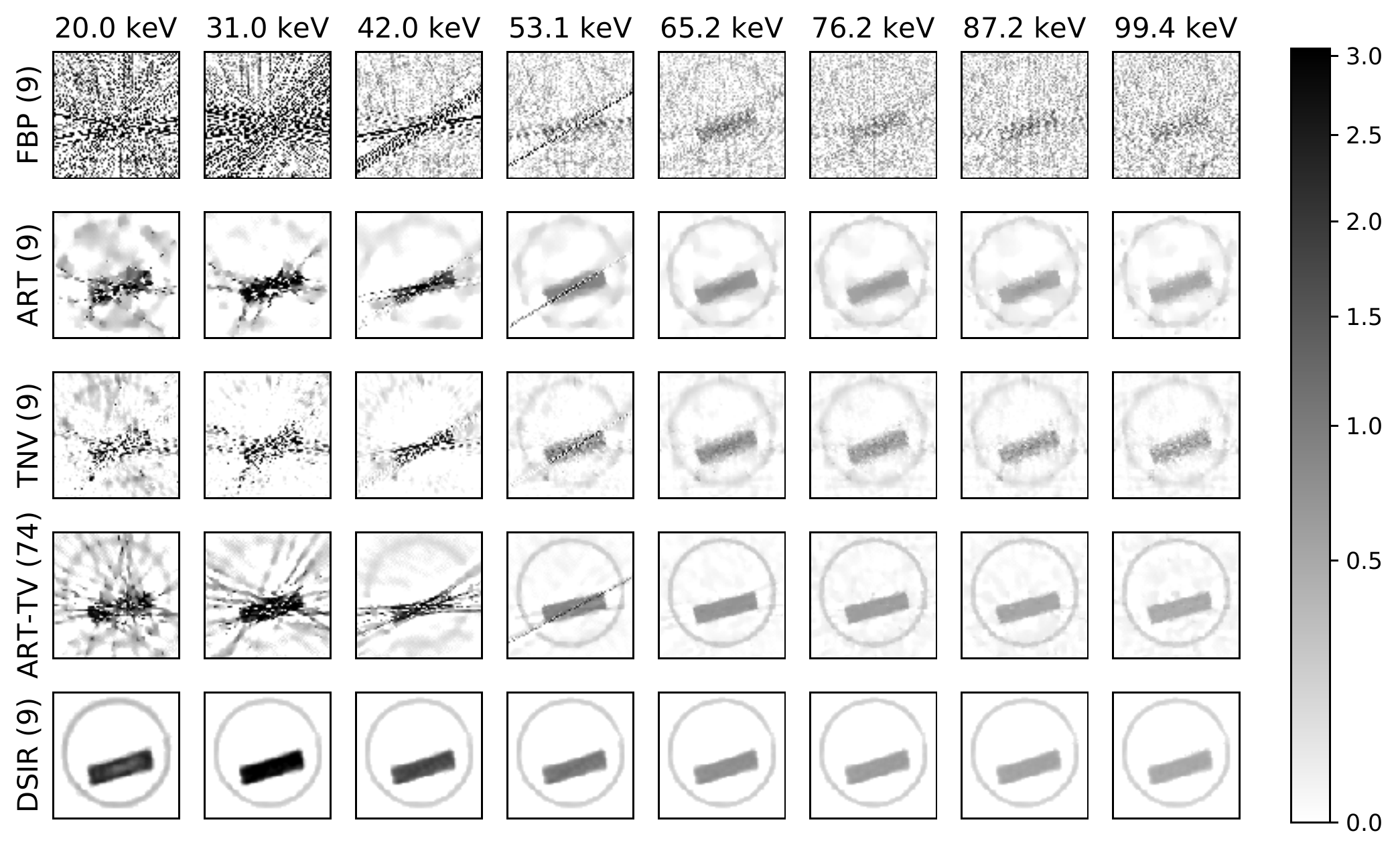}}
\subcaptionbox{}
{\includegraphics[width=0.50\textwidth]{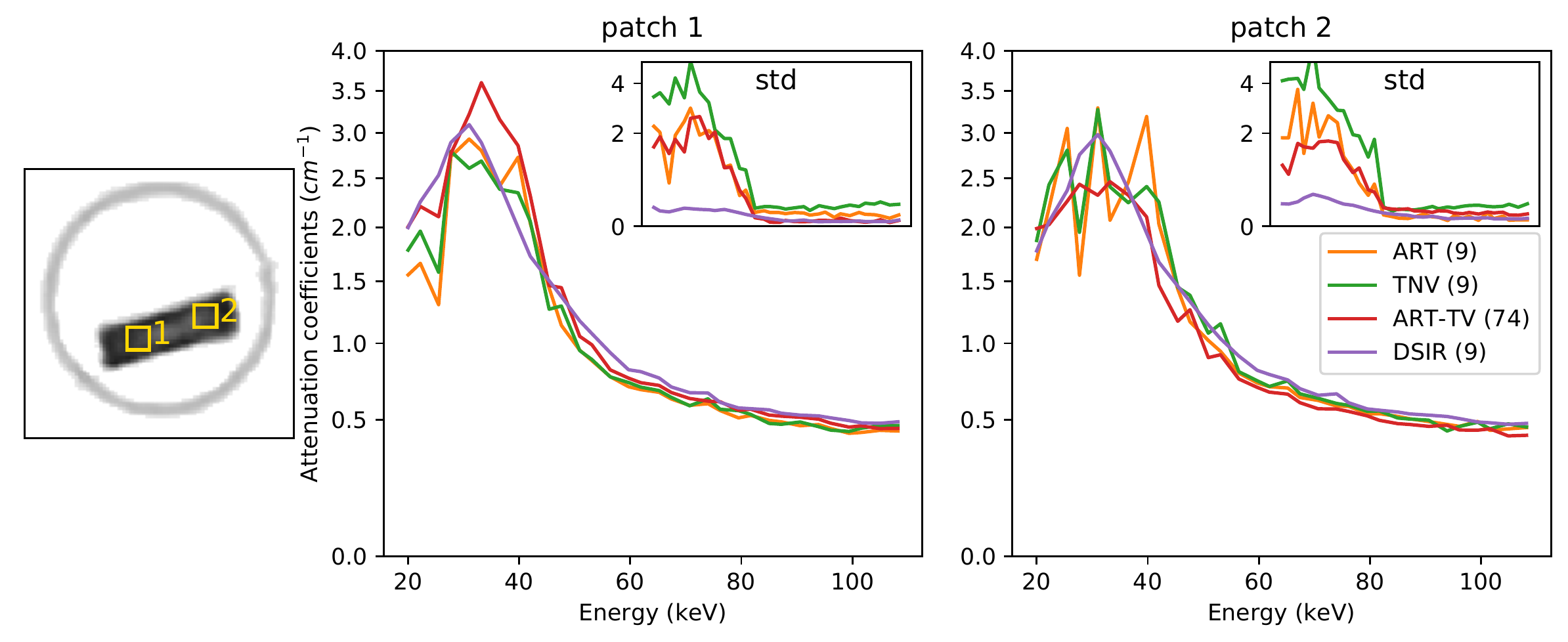}}
\subcaptionbox{}
{\includegraphics[width=0.16\textwidth]{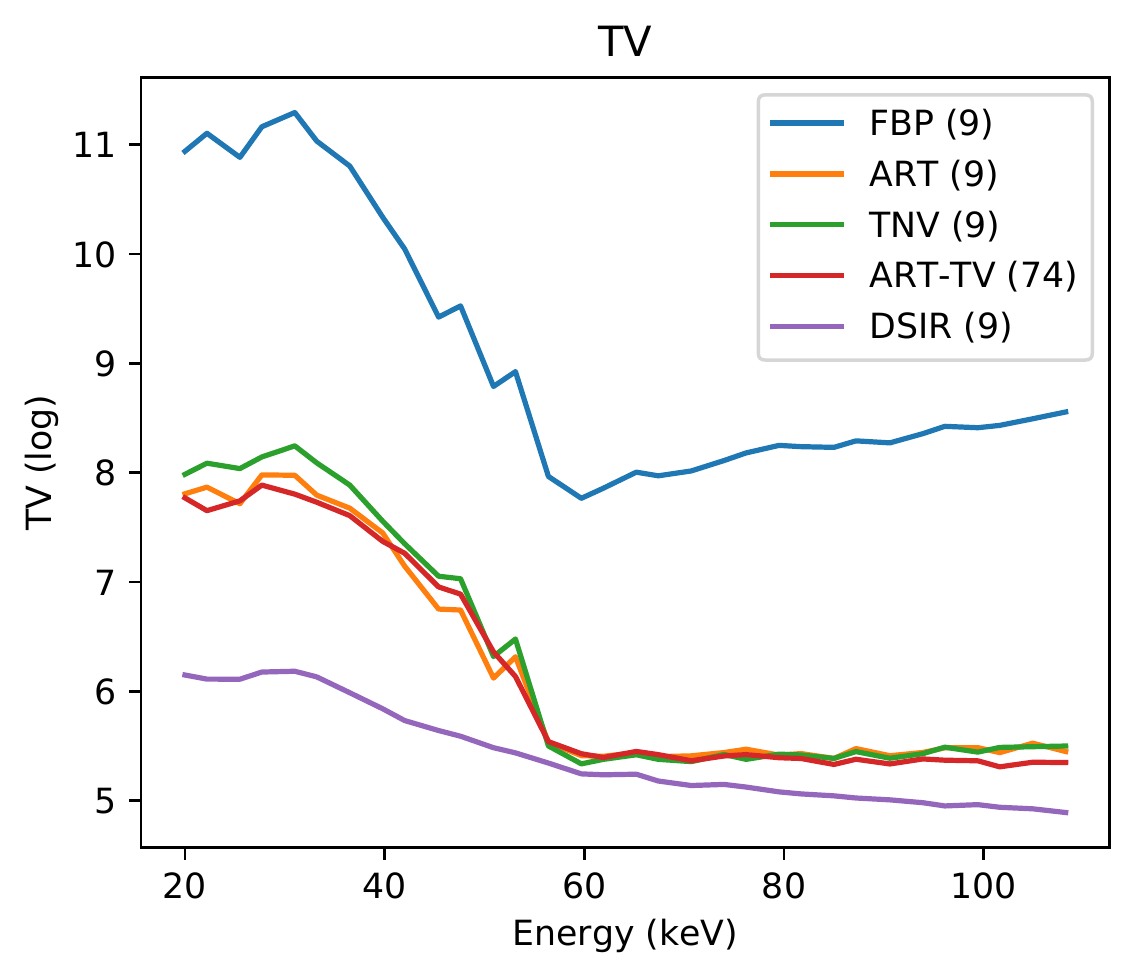}
\includegraphics[width=0.16\textwidth]{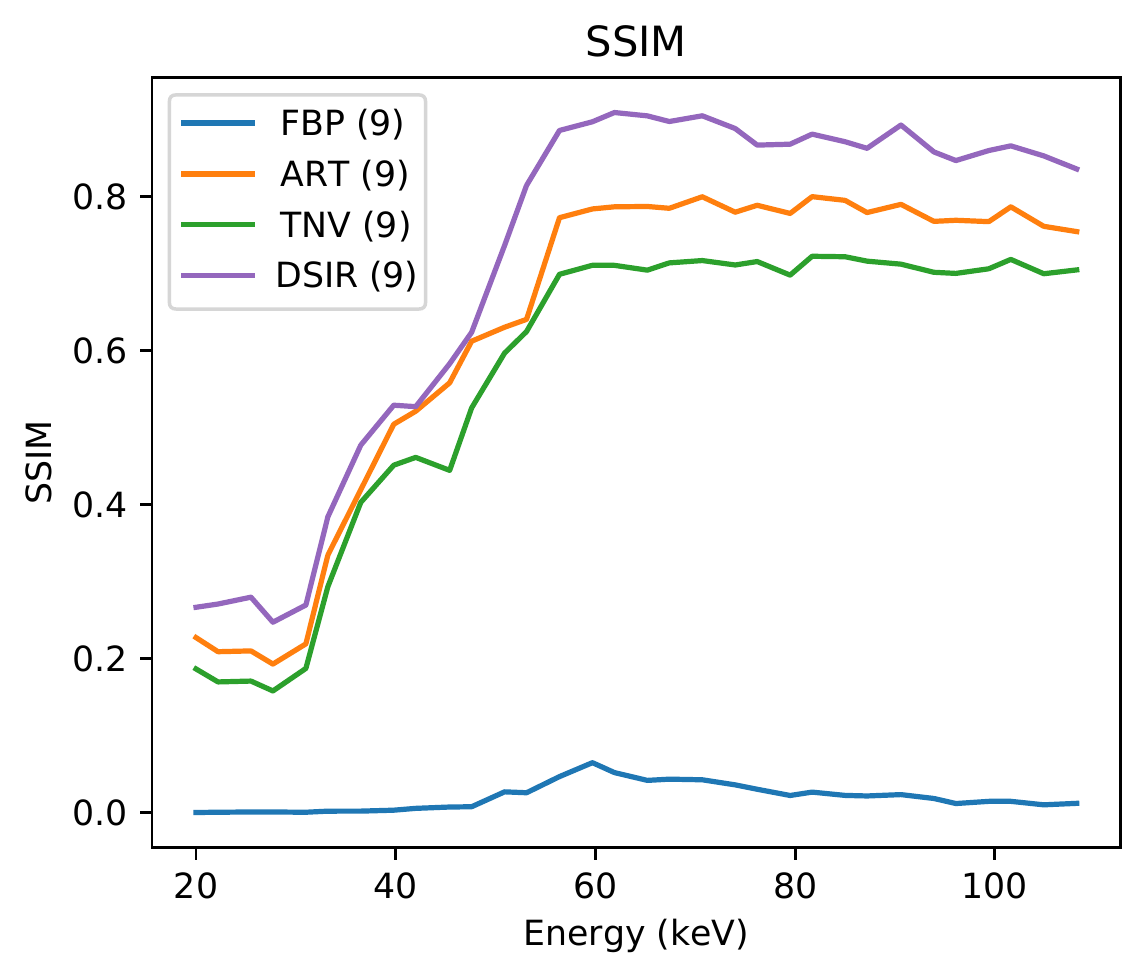}
\includegraphics[width=0.16\textwidth]{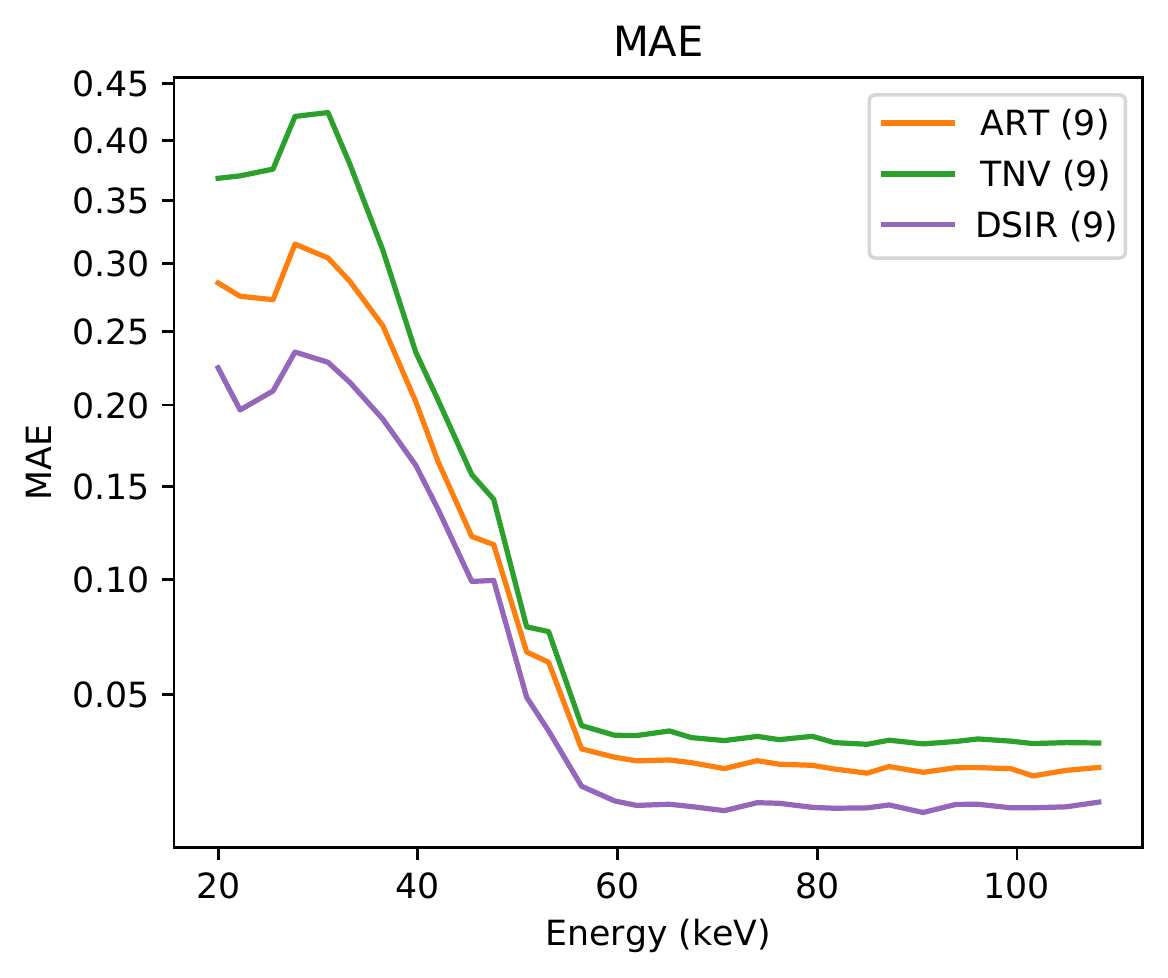}}

\caption{\label{fig:metal_art} Robustness to metal artifacts. The slice contains a large metal object causing an extreme case of metal artifacts. The slice is from the large rectangular shape in Volume I shown in \fig\ref{fig:volums}}
\end{figure}

\fig\ref{fig:metal_art} shows a comparison similar to the one presented in \fig\ref{fig:ours_vs_sot} but with a slice containing a large object made of Aluminum. \gls{CT} reconstruction of a sample containing a highly-absorbing material such as Aluminum is prone to streaks. In spectral data, the streaks appear predominantly at the lower end of the spectrum as we can see in the figure. This phenomenon, referred to as metal artifacts, is a common issue in single-energy \gls{CT} reconstruction. There is a rich literature of methods dedicated to address this issues \cite{gjesteby2016metal,wellenberg2018metal}. 

It is assumed that the use of spectral detectors makes it easier to tackle metal artifacts since the high-energy photons, with higher penetration, are isolated from the low-energy photons. This can be seen in the figure where higher reconstruction quality is obtained at higher energy channels. Metal artifacts reduction methods developed for spectral CT exploit this aspect \cite{schwahofer2015application,Nasirudin2015}.

Despite being an extreme case of metal artifacts induced by a relatively large metal object, the figure shows that our model is able to remove the streaks and produce high quality images. This further indicates that the model learned to exploit the high-SNR channels to compensate for the low-SNR channels. 


\subsection{Test set quantification}

In the subsections above, we have studied the reconstruction quality of our approach on individual slices selected from our dataset. In Table \ref{fig:performance_summary}, we present a quantification summary of the whole test set. The set is composed of 4 volumetric images (shown in \fig \ref{fig:volums}), each containing between 150 and 500 slices and with a total of 1175 slices. The table shows that our approach consistently performs the best. 


\begin{table*}[!tb]
\centering
\scriptsize
\begin{tabular}{l|ccc|ccc|ccc|ccc|c}
 & \multicolumn{3}{c|}{FBP (9)}  & \multicolumn{3}{c|}{ART-TV (9)} & \multicolumn{3}{c|}{TNV (9)} & \multicolumn{3}{c|}{DSIR (9)} & ART-TV (74) \\
 &
  \multicolumn{1}{c}{\cellcolor[HTML]{EFEFEF}TV} &
  \multicolumn{1}{c}{\cellcolor[HTML]{C0C0C0}SSIM} &
  \multicolumn{1}{c|}{\cellcolor[HTML]{9B9B9B}MAE} &
  \multicolumn{1}{c}{\cellcolor[HTML]{EFEFEF}TV} &
  \multicolumn{1}{c}{\cellcolor[HTML]{C0C0C0}SSIM} &
  \multicolumn{1}{c|}{\cellcolor[HTML]{9B9B9B}MAE} &
  \multicolumn{1}{c}{\cellcolor[HTML]{EFEFEF}TV} &
  \multicolumn{1}{c}{\cellcolor[HTML]{C0C0C0}SSIM} &
  \multicolumn{1}{c|}{\cellcolor[HTML]{9B9B9B}MAE} &
  \multicolumn{1}{c}{\cellcolor[HTML]{EFEFEF}TV} &
  \multicolumn{1}{c}{\cellcolor[HTML]{C0C0C0}SSIM} &
  \multicolumn{1}{c}{\cellcolor[HTML]{9B9B9B}MAE} &
  \multicolumn{1}{|c}{\cellcolor[HTML]{EFEFEF}TV} \\
\hline
Volume I &
  \cellcolor[HTML]{EFEFEF}11.83 &
  \cellcolor[HTML]{C0C0C0}0.04 &
  \cellcolor[HTML]{9B9B9B}1.873 &
  \cellcolor[HTML]{EFEFEF}8.74 &
  \cellcolor[HTML]{C0C0C0}0.73 &
  \cellcolor[HTML]{9B9B9B}0.102 &
  \cellcolor[HTML]{EFEFEF}9.31 &
  \cellcolor[HTML]{C0C0C0}0.67 &
  \cellcolor[HTML]{9B9B9B}0.191 &
  \cellcolor[HTML]{EFEFEF}7.75 &
  \cellcolor[HTML]{C0C0C0}0.84 &
  \cellcolor[HTML]{9B9B9B}0.073 &
  \cellcolor[HTML]{EFEFEF}8.55 \\
Volume II &
  \cellcolor[HTML]{EFEFEF}11.45 &
  \cellcolor[HTML]{C0C0C0}0.02 &
  \cellcolor[HTML]{9B9B9B}1.465 &
  \cellcolor[HTML]{EFEFEF}8.30 &
  \cellcolor[HTML]{C0C0C0}0.72 &
  \cellcolor[HTML]{9B9B9B}0.074 &
  \cellcolor[HTML]{EFEFEF}8.91 &
  \cellcolor[HTML]{C0C0C0}0.72 &
  \cellcolor[HTML]{9B9B9B}0.140 &
  \cellcolor[HTML]{EFEFEF}6.92 &
  \cellcolor[HTML]{C0C0C0}0.82 &
  \cellcolor[HTML]{9B9B9B}0.052 &
  \cellcolor[HTML]{EFEFEF}8.05 \\
Volume III &
  \cellcolor[HTML]{EFEFEF}11.67 &
  \cellcolor[HTML]{C0C0C0}0.02 &
  \cellcolor[HTML]{9B9B9B}1.379 &
  \cellcolor[HTML]{EFEFEF}8.52 &
  \cellcolor[HTML]{C0C0C0}0.72 &
  \cellcolor[HTML]{9B9B9B}0.074 &
  \cellcolor[HTML]{EFEFEF}8.91 &
  \cellcolor[HTML]{C0C0C0}0.66 &
  \cellcolor[HTML]{9B9B9B}0.120 &
  \cellcolor[HTML]{EFEFEF}7.46 &
  \cellcolor[HTML]{C0C0C0}0.83 &
  \cellcolor[HTML]{9B9B9B}0.051 &
  \cellcolor[HTML]{EFEFEF}8.37 \\
Volume IV &
  \cellcolor[HTML]{EFEFEF}11.55 &
  \cellcolor[HTML]{C0C0C0}0.01 &
  \cellcolor[HTML]{9B9B9B}1.631 &
  \cellcolor[HTML]{EFEFEF}8.40 &
  \cellcolor[HTML]{C0C0C0}0.77 &
  \cellcolor[HTML]{9B9B9B}0.077 &
  \cellcolor[HTML]{EFEFEF}9.04 &
  \cellcolor[HTML]{C0C0C0}0.78 &
  \cellcolor[HTML]{9B9B9B}0.151 &
  \cellcolor[HTML]{EFEFEF}6.85 &
  \cellcolor[HTML]{C0C0C0}0.80 &
  \cellcolor[HTML]{9B9B9B}0.059 &
  \cellcolor[HTML]{EFEFEF}8.08 \\
  \hline
  mean &
  \cellcolor[HTML]{EFEFEF}11.54 &
  \cellcolor[HTML]{C0C0C0}0.02 &
  \cellcolor[HTML]{9B9B9B}1.587 &
  \cellcolor[HTML]{EFEFEF}8.45 &
  \cellcolor[HTML]{C0C0C0}0.72 &
  \cellcolor[HTML]{9B9B9B}0.082 &
  \cellcolor[HTML]{EFEFEF}8.89 &
  \cellcolor[HTML]{C0C0C0}0.69 &
  \cellcolor[HTML]{9B9B9B}0.151 &
  \cellcolor[HTML]{EFEFEF}7.29 &
  \cellcolor[HTML]{C0C0C0}0.81 &
  \cellcolor[HTML]{9B9B9B}0.059 &
  \cellcolor[HTML]{EFEFEF}8.29\\
 &         &         &         &          &          &          &         &         &         &        &        &        &            
\end{tabular}
\caption{\label{fig:performance_summary} Comparison on the test set. The set contains four volumetric (multi-slice) images. The volume mean of each metric is shown. The overall mean is shown at the bottom.
}

\end{table*}

\newcommand\subWidth{0.07\textwidth}
\begin{figure}
\centering

\begin{tabular}{ccccc}
\hline
 ART-TV (9) & TNV (9) & DSIR (9) &  Ground truth \\
 \hline
  \\
 \includegraphics[width=\subWidth]{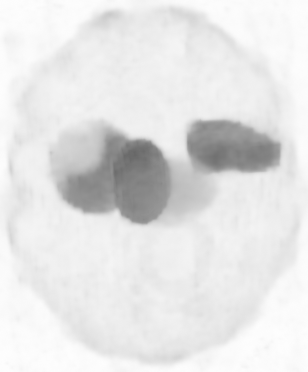}  & \includegraphics[width=\subWidth]{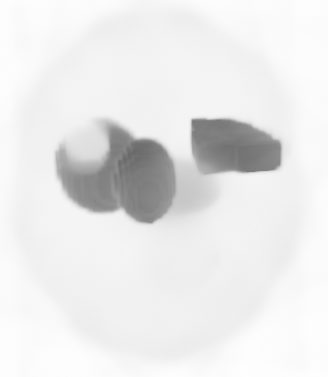}  &  \includegraphics[width=\subWidth]{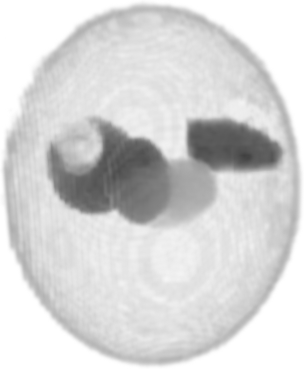} &  \includegraphics[width=\subWidth]{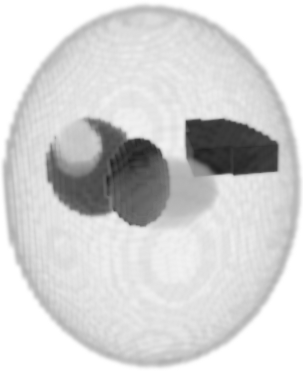} \\
\multicolumn{4}{c}{Synthetic Sample}  \\
 \\
 \hline
 ART-TV (9) & TNV (9) & DSIR (9) &  ART-TV (74) \\
 \hline
  \\
 \includegraphics[width=\subWidth]{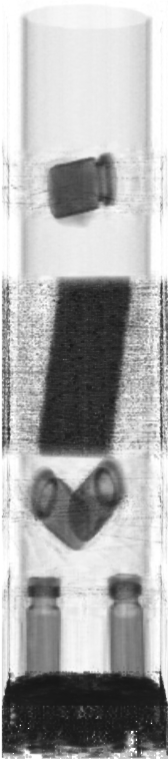}  & \includegraphics[width=\subWidth]{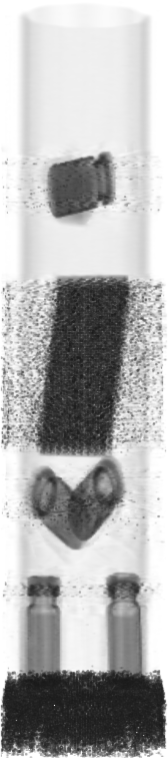}  &  \includegraphics[width=\subWidth]{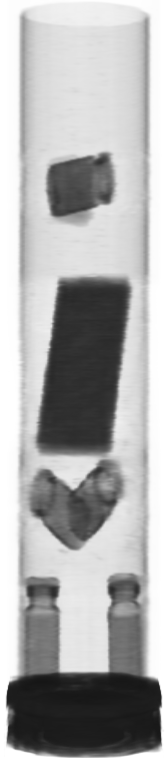} &  \includegraphics[width=\subWidth]{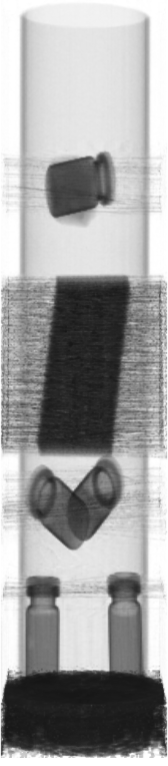} \\
\multicolumn{4}{c}{Volume I (Sample\_23012018)}  \\
 \\

 \includegraphics[width=\subWidth]{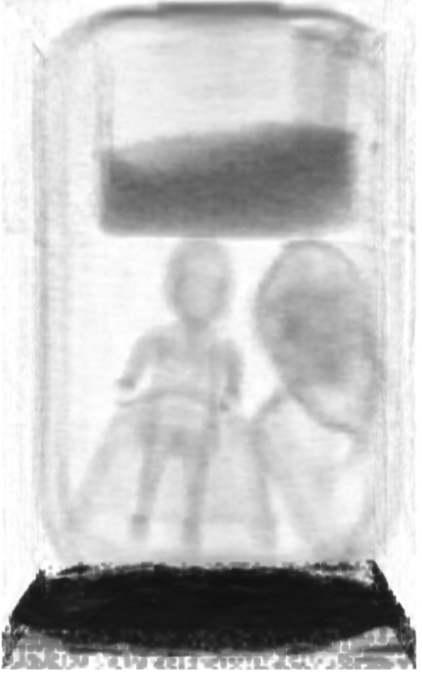}  & \includegraphics[width=\subWidth]{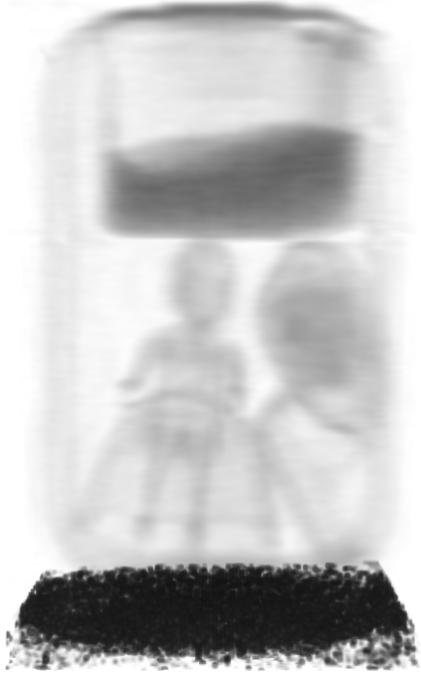}  &  \includegraphics[width=\subWidth]{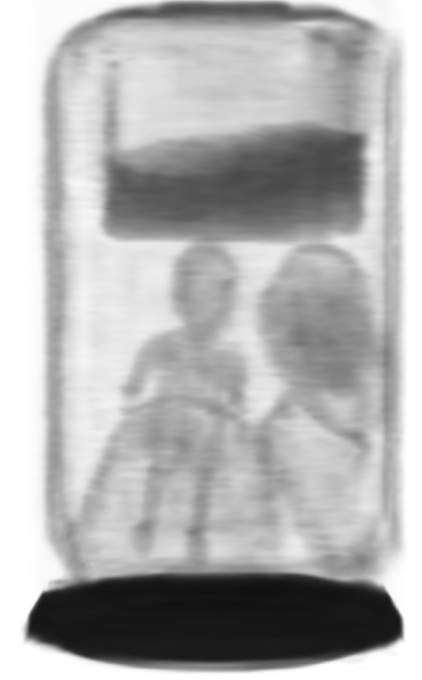} &  \includegraphics[width=\subWidth]{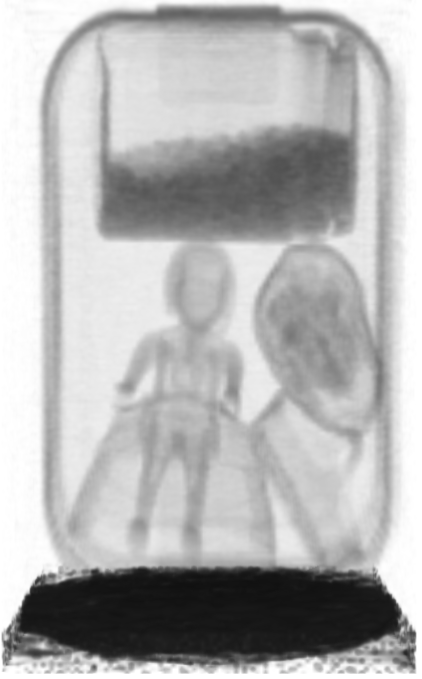} \\
\multicolumn{4}{c}{Volume II (181128\_Phantom\_6\_002)} \\
 \\

 \includegraphics[width=\subWidth]{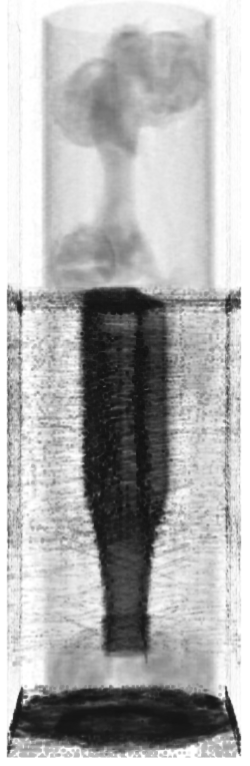}  & \includegraphics[width=\subWidth]{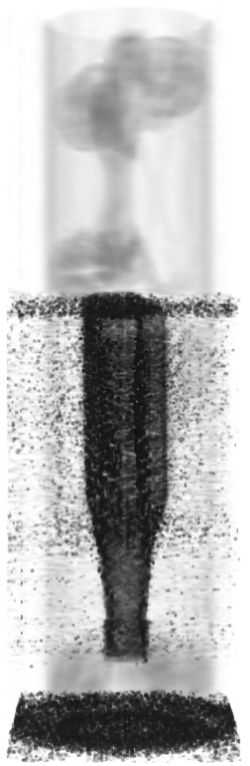}  &  \includegraphics[width=\subWidth]{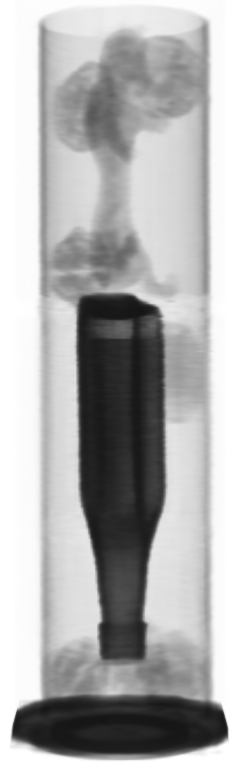} &  \includegraphics[width=\subWidth]{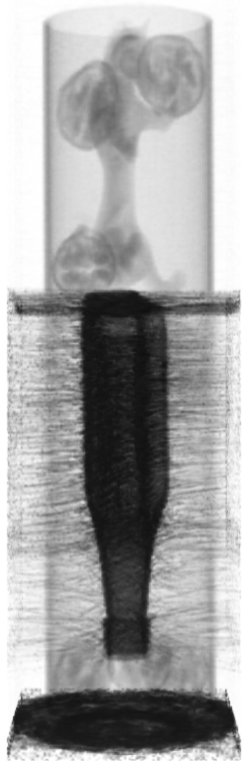} \\
\multicolumn{4}{c}{Volume III (181128\_Phantom\_2\_005)} \\
 \\

 \includegraphics[width=\subWidth]{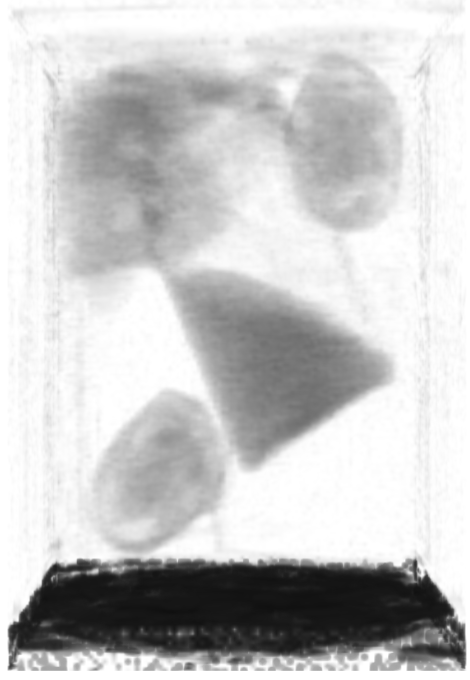}  & \includegraphics[width=\subWidth]{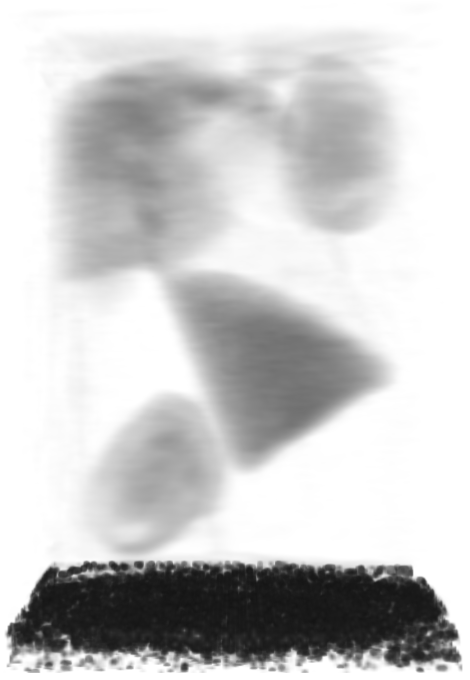}  &  \includegraphics[width=\subWidth]{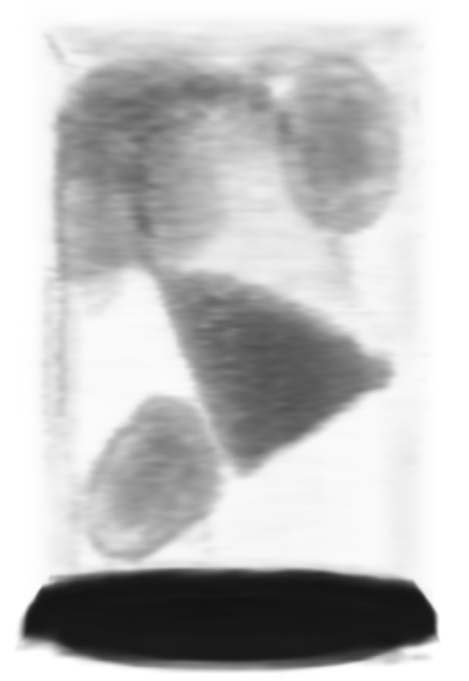} &  \includegraphics[width=\subWidth]{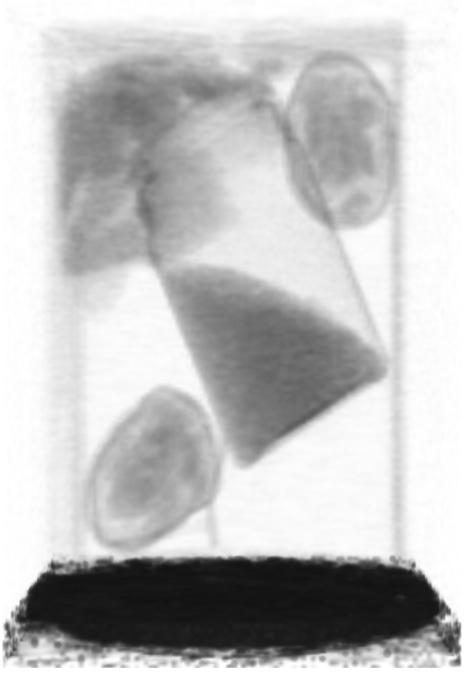} \\
\multicolumn{4}{c}{Volume IV (181128\_Phantom\_1\_003)}

\end{tabular}

\caption{\label{fig:volums} Renderings of the four volumes (channel $53.1~keV$) composing the test set. Each volume is reconstructed slice by slice. The volume's name in the MUSIC dataset is shown between parentheses. }
\end{figure}


\subsection{Computation time}

In this subsection, we present run-time comparison of our approach with respect to the other state-of-the-art methods we consider in this paper. This comparison supports the computational complexity analysis discussed in \sect \ref{sec:arch:complexity}. The main challenge in conducting such comparisons is that implementations vary significantly from one method to another. In order to provide fair comparison, we consider mainly the CPU implementations written in C++. Furthermore, we force the execution to utilize only one CPU core to limit parallelism. Moreover, for implementations with Python interface, we measure the execution time of the primitive function invoking a C++ function. Because we don't have an optimized C++ implementation for the TNV method, we exclude it from this comparison.


Table \ref{fig:evaluation_time} summarizes the execution time of the methods in comparison using both their CPU and GPU implementations. The evaluations are performed on a machine running a Linux system and equipped with an Intel Xeon (3.0 GHz) E5-2687W v4 x 48 CPU and an NVIDIA GeForce GTX 1080 GPU (20 streaming multiprocessors, each with 128 CUDA-cores at 1.6 GHz base clock). 

The table shows that our approach is significantly faster than \gls{ART-TV} in both the CPU and the GPU implementations. When comparing the speedup achieved with GPU implementation however, we find that we gain about 37 times speedup with ART-TV and about 7 times speedup with our approach. This disparity in speed boost can be attributed to the fact that for \gls{ART-TV}, we use an implementation optimized for performance and data access patterns whereas for \gls{CNN}, we use off-the-shelf packages made for generic prototyping.

One aspect to note is that the evaluation time for the CNN part of our approach (i.e., with 32 channels) is 6.4 ms while it is 4.32 ms for a single-channel version of the network (discussed in \sect \ref{sec:result_jointVSsingle}). This highlights the speed advantage of using our spectral CNN, as the internal layers are shared between the channel. Furthermore, in our computational complexity analysis in \ref{sec:arch:complexity}, we estimated that the complexity of our approach was dominated by the \gls{CNN} part rather than the \gls{FBP}. The table shows that with CPU implementations, the \gls{FBP} part requires less time than the \gls{CNN} part, which supports our estimation.


\begin{table}[!tp]
\centering
\scriptsize

\begin{tabular}{l|r|r|r|r|}
\cline{2-5}
                               &  \multicolumn{2}{c|}{CPU} & \multicolumn{2}{|c|}{GPU} \\ 
                               \cline{2-5}
                                            &  1 channel &  32 channels & 1 channel & 32 channels      \\ \hline
\multicolumn{1}{|l|}{ART-TV (9)}            & 25.70      &  \textit{822.35}      & 0.69      &        \textit{22.00}     \\ \hline
\multicolumn{1}{|l|}{FBP (9)}               & 0.14       &  \textit{4.50}        & 0.03      &        \textit{0.90}      \\ \hline
\multicolumn{1}{|l|}{CNN (9)}               & \textit{0.20}       &  6.40        & \textit{0.02}      &        0.60      \\ \hline
\multicolumn{1}{|l|}{CNN\_sc (9)}            & 4.32       &  \textit{138.30}      & 3.35      &      \textit{77.10}       \\ \hline
\multicolumn{1}{|l|}{DSIR (9) [FBP+CNN]}        & \textit{0.34}       &  \textit{10.90}        & \textit{0.05}      &      \textit{1.50}        \\ \hline
\end{tabular}

\caption{\label{fig:evaluation_time} Comparison of the evaluation time (ms). We show the time needed to process a single channel (1 channel) and to process the whole spectrum (32 channel). 
Italicized numbers are estimates (not actual measurements) computed to facilitate the comparison. CNN\_sc is a single-channel version of our network (discussed in \sect \ref{sec:result_jointVSsingle}).
}
\end{table}

\section{Conclusion}
In this paper, we presented an approach for reconstructing spectral CT from sparse-view projections. The approach is based on multi-channel U-Net architecture trained to remove noise and artifacts from images initially reconstructed by \gls{FBP}. The network processes the spectral channels jointly using shared convolutions, making our approach significantly faster than the iterative reconstruction methods. Moreover, the results show that our approach outperforms the state of the art methods in terms of both the spatial and spectral reconstruction quality.

The results obtained in this paper demonstrates the importance of coupling the spectral channels for sparse-view reconstruction. First, our results indicate that channel coupling allows the network to circumvent the low \gls{SNR} provided by spectral detectors. Second, our results strongly suggest that channel coupling enables the network to utilize healthier channels to compensate for artifacts affecting other channels. This was demonstrated by the robustness of the network to channels with significantly higher noise levels and also to metal artifacts, which affect primarily the low-energy channels. This robustness is driven by the spectral consistency the network learns to maintain. The coupling of channels in such a way allows us to maximize the gain of using spectral detectors.




Compared with the iterative methods, our approach provides a strong alternative that is faster and more effective for sparse-view reconstruction. Applying \gls{CNN}, however, raises the question of generalization. Even though generalization is beyond the scope of this paper, we applied data augmentation to mitigate potential over-fitting and by using a dedicated test set, it is easier to detect signs of over-fitting. That said, the test set resembles the training set with respect to several aspects such as acquisition settings, object geometries, and materials. It is therefore difficult to draw final conclusion on the network generalization properties, especially regarding those aspects. Th aspects will be further explored in future work. 


     

\section*{Acknowledgment}

The authors would like to thank Sean Meyer, Qiyuan Liang, Yijie Jin, Mustafa Al-Hamdani, and Ragavan Pathanchalinathan for their early efforts within student projects on the topic. The authors would also like to thank Jakeoung Koo for providing an implementation of the \gls{TNV} method used for comparison. This research is supported by Innovation Fund Denmark (project  10437) and the EIC FTI program (project 853720). 

\bibliographystyle{IEEEtran}
\bibliography{SpectralCTNN}

\vskip -2\baselineskip plus -1fil

\begin{IEEEbiography}
[{\includegraphics[width=1in,height=1.25in,clip,keepaspectratio]{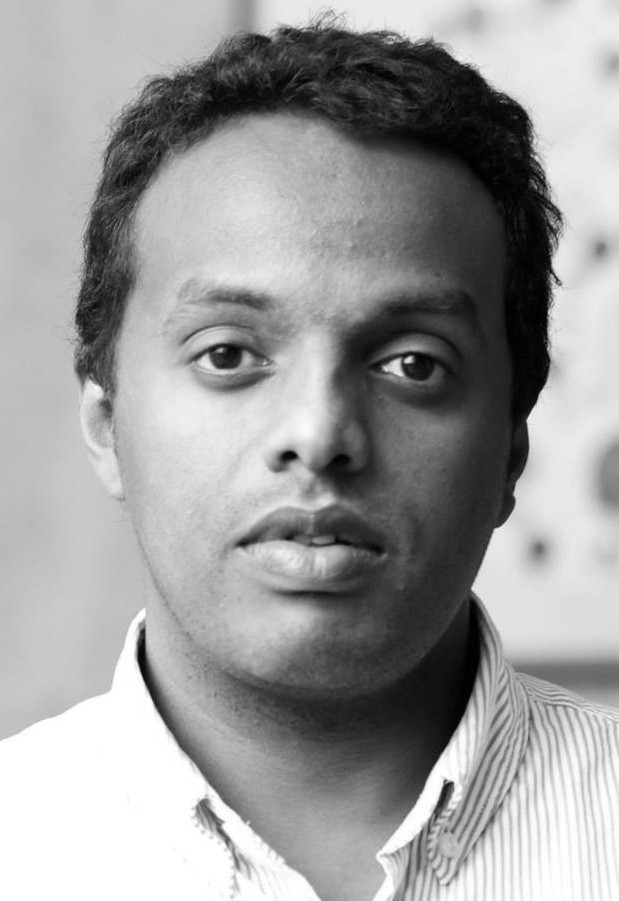}}]{Wail Mustafa}
received his B.Sc degree in electrical engineering from University of Khartoum, Sudan, in 2006, his M.Sc. degree in electrical engineering (signal processing) from Blekinge Institute of Technology, Sweden, in 2010, and his PhD degree in robotics (computer vision) from University of Southern Denmark in 2015. Since 2017, he has been a postdoctoral researcher at The Technical University of Denmark (DTU). His main research interests are computer vision and machine learning for cognitive and autonomous systems.
\end{IEEEbiography}

\vskip -2\baselineskip plus -1fil

\begin{IEEEbiography}[{\includegraphics[width=1in,height=1.25in,clip,keepaspectratio]{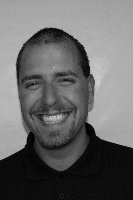}}]{Christian Kehl}
is a Postdoc at Utrecht University (Dept. of Physics) since 2020. Christian received his Ph.D. in Mathematics and Natural Sciences in 2017 from the University of Bergen (NO). He had former research affiliations with TU Delft (NL), University of Amsterdam (NL), Uni Research (NO) and Aix-Marseille Universit\'{e} (FR), DTU Compute (DK) and Fraunhofer Institute (DE). His research interests are high-performance computing, image analysis, scientific visualisation, and discrete geometry computing.
\end{IEEEbiography}

\vskip -2\baselineskip plus -1fil

\begin{IEEEbiography}[{\includegraphics[width=1in,height=1.25in,clip,keepaspectratio]{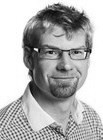}}]{Ulrik Lund Olsen}
is a Sr. research engineer at DTU physics since 2013. Funded by EC through H2020 to develop applications using high flux multispectral x-ray detection technology and currently project coordinator on XSPERINSE, a multidisciplinary effort made to reduce the human operator involvement by 50\% for checked-in luggage. He was previously employed at Ris\o\ National Laboratory working with developments of x-ray sensors.
\end{IEEEbiography}

\vskip -2\baselineskip plus -1fil

\begin{IEEEbiography}[{\includegraphics[width=1in,height=1.25in,clip,keepaspectratio]{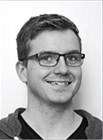}}]{S{\o}ren~Kimmer~Schou~Gregersen}
is an assistant professor at the Technical University of Denmark (Kns. Lyngby, Denmark). He received his Ph.D. in physics and nanotechnology in 2017 and has since studied computer vision at the Technical University of Denmark. His research interests are optical 3D scanning using structured light and modelling interactions between light and matter.
\end{IEEEbiography}

\vskip -2\baselineskip plus -1fil

\begin{IEEEbiography}[{\includegraphics[width=1in,height=1.25in,clip,keepaspectratio]{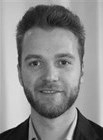}}]{David~Malmgren-Hansen}
is a Postdoc at the Technical University of Denmark Department of Applied Mathematics and Computer Science. His research field is centered around developing and applying Deep Learning in different fields and has worked extensively in with various types of problems within remote sensing.
He holds a MS degree in Electrical Engineering and a PhD in Computer Science from the the Technical University of Denmark.
\end{IEEEbiography}

\vskip -2\baselineskip plus -1fil

\begin{IEEEbiography}[{\includegraphics[width=1in,height=1.25in,clip,keepaspectratio]{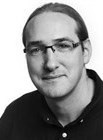}}]{Jan Kehres}
is a research engineer at Technical University of Denmark (DTU) Physics, Kongens Lyngby, Denmark. He received his PhD in operando investigation of catalyst nanoparticles using X-ray scattering and continued in this field of research as a postdoctoral researcher. His current field of research is the application of energy-dispersive detectors for X-ray scattering and advanced imaging modalities with a focus on material identification of illicit materials for security applications.
\end{IEEEbiography}

\vskip -2\baselineskip plus -1fil

\begin{IEEEbiography}[{\includegraphics[width=1in,height=1.25in,clip,keepaspectratio]{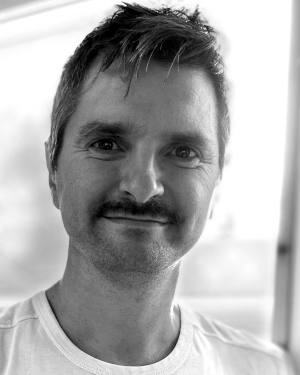}}]{Anders Bjorholm Dahl}
is a professor in 3D image analysis at the Technical University of Denmark from where he also received his PhD degree in 2009. His research covers computer vision and 3D image analysis including tomographic reconstruction techniques, segmentation methods, graphical methods, and learning based methods from weakly or partly labeled data. These methods are used in biomedical applications, materials science applications and others, and a specific focus is within analysis of micro-structure in images from X-ray CT and neutron scattering.
\end{IEEEbiography}

\end{document}